\documentclass[10pt,journal,compsoc,onecolumn]{IEEEtran}
\usepackage{ifpdf}
\ifCLASSOPTIONcompsoc
  \usepackage[nocompress]{cite}
\else
  \usepackage{cite}
\fi
\ifCLASSINFOpdf
  \usepackage[pdftex]{graphicx}
  \DeclareGraphicsExtensions{.pdf,.jpeg,.jpg,.png}
\else
  \usepackage[dvips]{graphicx}
  \DeclareGraphicsExtensions{.eps}
\fi
\ifpdf
\else
	\usepackage{psfrag}
\fi

\usepackage{amsmath}
\usepackage{array}
\ifCLASSOPTIONcompsoc
  \usepackage[caption=false,font=footnotesize,labelfont=sf,textfont=sf]{subfig}
\else
  \usepackage[caption=false,font=footnotesize]{subfig}
\fi
\usepackage{dblfloatfix}
\usepackage{url}
\usepackage{threeparttable}

\newcommand{\arXiv}[2]{#1}

\newcommand{\TCBB}[2]{#1}		
\TCBB{
\newcommand{\SkipTCBBToMerge}[1]{#1}	
\newcommand{\SkipNonTCBBToMerge}[1]{}	
}{
\newcommand{\SkipTCBBToMerge}[1]{}	
\newcommand{\SkipNonTCBBToMerge}[1]{#1}	
}

\ifdefined\SUPPLEMENT
\else
\newcommand{\SUPPLEMENT}[1]{}
\fi

\newcommand{\SkipSupplToMerge}[1]{}
\SUPPLEMENT{
\renewcommand{\SkipSupplToMerge}[1]{#1}
}

\ifdefined\TEXT
\else
\newcommand{\TEXT}[1]{#1}
\fi

\newcommand{\SkipTextToMerge}[1]{}
\TEXT{
\renewcommand{\SkipTextToMerge}[1]{#1}
}

\TCBB{
\usepackage{ifthen,calc}
\usepackage{xifthen}
\usepackage{ifthenx}

\ifdefined\Skip
\else
\newcommand{\Skip}[1]{}
\fi

\usepackage{color}
\usepackage{times}
\usepackage{txfonts}
\usepackage{bm}

\usepackage[hidelinks]{hyperref}

\ifdefined\TCBB
\else
\newcommand{\TCBB}[2]{#2}
\fi
\ifdefined\arXiv
\else
\newcommand{\arXiv}[2]{#2}
\fi
\ifdefined\BMC
\else
\newcommand{\BMC}[2]{#2}
\fi

\ifdefined\Skip
\else
\newcommand{\Skip}[1]{}                 
\fi

\newcommand{\script}[1]{{\mbox{\scriptsize #1}}}

\newcommand{\CITE}[1]{\cite{#1}}
\newcommand{\VEC}[1]{\text{\boldmath{$#1$}}}
\newcommand{\VECS}[1]{\text{\boldmath{$#1$}}}
\newcommand{\Eqno}[1]{{#1}}
\newcommand{\Eq}[1]{Eq. \Eqno{#1}}
\newcommand{\Eqs}[1]{Eqs. {#1}}

\newcommand{\Fig}[1]{Fig. {#1}}
\newcommand{\Figs}[1]{Figs. {#1}}
\newcommand{\SFig}[1]{Fig. {#1}}
\newcommand{\SFigs}[1]{Figs. {#1}}
\newcommand{\Table}[1]{Table {#1}}

\newcommand{\SkipFigure}[1]{}
\newcommand{\FigureInText}[1]{}

\newcommand{\FigureInLegends}[1]{}

\newcommand{\TableInText}[1]{}
\newcommand{\TableLegends}[1]{}

\newcommand{\EQ}{eq}
\newcommand{\TBL}{tbl}
\newcommand{\FIG}{fig}
\newcommand{\SECTION}[1]{\section{#1}}
\newcommand{\SUBSECTION}[1]{\subsection{#1}}

\newcommand{\TEXTBF}[1]{#1}

\ifdefined\EQ
\renewcommand{\EQ}{eq}
\else
\newcommand{\EQ}{eq}
\fi
\ifdefined\TBL
\renewcommand{\TBL}{tbl}
\else
\newcommand{\TBL}{tbl}
\fi
\ifdefined\FIG
\renewcommand{\FIG}{fig}
\else
\newcommand{\FIG}{fig}
\fi

\newcommand{\Detail}[1]{}

\ifdefined\text
\else
\newcommand{\text}[1]{\textrm{#1}}
\fi

\renewcommand{\VEC}[1]{\text{\boldmath{$#1$}}} 
\renewcommand{\VECS}[1]{\text{\boldmath{$#1$}}}

\renewcommand{\SFig}[1]{\Fig{#1}}
\renewcommand{\SFigs}[1]{\Figs{#1}}

\newcommand{\SupplementaryBibliography}[1]{#1}

\newcommand{\TextMaterial}[1]{#1}
\TextMaterial{
\renewcommand{\SupplementaryBibliography}[1]{}
}

\renewcommand{\FigureInText}[1]{#1}
\renewcommand{\TableInText}[1]{#1}

\newcommand{\SkipFigsInText}[1]{}
\FigureInText{
\renewcommand{\SkipFigsInText}[1]{#1}
}% FigureInText
\newcommand{\SkipTablesInText}[1]{}
\TableInText{
\renewcommand{\SkipTablesInText}[1]{#1}
}% TableInText

}{

}% TCBB

\newcommand{\NAG}[2]{#1}	% NAG is included.
\newcommand{\FiguresWithoutCaption}[1]{}
\newcommand{\withSup}[2]{#1}
\FiguresWithoutCaption{
\renewcommand{\withSup}[2]{#2}
}

\begin{document}
%\title{Bare Demo of IEEEtran.cls for\\ IEEE Computer Society Journals}
\title{
Boltzmann Machine Learning with a Parallel, 
\newline
Persistent Markov chain Monte Carlo method
\newline
for Estimating Evolutionary Fields and Couplings 
\newline
from a Protein Multiple Sequence Alignment
% End of title.tex
}
% \IEEEcompsocthanksitem becomes a line break with indentation. This

\author{Sanzo~Miyazawa%,~\IEEEmembership{Member,~BPSJ}
\IEEEcompsocitemizethanks{\IEEEcompsocthanksitem	%\protect\\
Email: sanzo.miyazawa@gmail.com
}%
%\thanks{Manuscript received 20 Sept. 2019; revised 8 Apr. 2020; accepted 2 May 2020;
%DOI: 10.1109/TCBB.2020.2993232 ; The corrections made afterward are indicated in red.}
}


%\markboth{IEEE/ACM Trans. Comput. Biol. Bioinform., 2020}%
\markboth{  }%
{Sanzo Miyazawa: Boltzmann machine learning for an inverse Potts problem}



\IEEEtitleabstractindextext{%
\begin{abstract}
%The abstract goes here.
%\input{abst_TCBB.tex}
% \input{abst.tex}
The inverse Potts problem 
for estimating evolutionary single-site fields and pairwise couplings in homologous protein sequences
from their single-site and pairwise amino acid frequencies
observed in their multiple sequence alignment
would be still one of useful methods
in the studies of protein structure and evolution.
Since the reproducibility of fields and couplings are the most important,
the Boltzmann machine method is employed here, although it is computationally
intensive. 
In order to reduce computational time required for the Boltzmann machine, 
parallel, persistent Markov chain Monte Carlo method is employed to estimate 
the single-site and pairwise marginal distributions in each learning step.
Also, stochastic gradient descent methods are used to reduce computational time for each learning.
Another problem is how to adjust
the values of hyperparameters; there are two regularization parameters
for evolutionary fields and couplings. 
The precision of contact residue pair prediction
is often used to adjust the hyperparameters.
However, it is not sensitive to these regularization parameters. 
Here, 
they are adjusted for the fields and couplings
to satisfy a specific condition that is appropriate for protein conformations.
This method has been applied to eight protein families.

% End of abst.tex
\end{abstract}

\begin{IEEEkeywords}
%Computer Society, IEEE, IEEEtran, journal, \LaTeX, paper, template.
%% from https://ieeecs-media.computer.org/assets/pdf/taxonomy.pdf
inverse Potts problem,
Boltzmann machine,
evolutionary fields and couplings,
protein multiple sequence alignment,
parallel persistent Markov chain Monte Carlo method,
learning schedule,
%hyperparameter adjustment
regularization parameter adjustment
\end{IEEEkeywords}}

\maketitle

\IEEEdisplaynontitleabstractindextext

% \fi
\IEEEpeerreviewmaketitle

\IEEEraisesectionheading{\section{Introduction}\label{sec:introduction}}


% local settings
\ifdefined\NAG

\withSup{
\newcommand{\SFigPFvModAdamvsAdamhJPiaCijab}{\SFig{\ref{sfig: PF00595_ModAdam_vs_Adam_hJ_PiaCijab}}}
\newcommand{\SFigPFvmaxJijcomparison}{\SFig{\ref{sfig: PF00595_maxJij_comparison}}}

\newcommand{\SFigPFvModAdamvsNAGhJPiaCijab}{\SFig{\ref{sfig: PF00595_ModAdam_vs_NAG_hJ_PiaCijab}}}

\newcommand{\SFigPFvModAdamvsRPROPLRhJPiaCijab}{\SFig{\ref{sfig: PF00595_ModAdam_vs_RPROP-LR_hJ_PiaCijab}}}
\newcommand{\SFigPFvmaxJijvsrpd}{\SFig{\ref{sfig: PF00595_maxJij_vs_rpd}}}
\newcommand{\SFigPFvhJcomparison}{\SFig{\ref{sfig: PF00595_hJ_comparison}}}
\newcommand{\SFigPFihJcomparison}{\SFig{\ref{sfig: PF00153_hJ_comparison}}}
\newcommand{\SFigsPFvPFihJcomparison}{\SFigs{\ref{sfig: PF00595_hJ_comparison} and \ref{sfig: PF00153_hJ_comparison}}}

\newcommand{\SFigsPFvPFiKL}{\SFig{\ref{sfig: learning_process_KL}}}

\newcommand{\SFigsPFvPFiPiaCijab}{\SFigs{\ref{fig: PF00595_PiaCijab} and \ref{fig: PF00153_PiaCijab}}}

\newcommand{\SFigPFvMCPiaCijab}{\SFig{\ref{sfig: PF00595MC_PiaCijab}}}
\newcommand{\SFigPFiMCPiaCijab}{\SFig{\ref{sfig: PF00153MC_PiaCijab}}}
\newcommand{\SFigsPFvMCPFiMCPiaCijab}{\SFigs{\ref{sfig: PF00595MC_PiaCijab} and \ref{sfig: PF00153MC_PiaCijab}}}

}{

\TCBB{

\newcommand{\SFigPFvModAdamvsAdamhJPiaCijab}{\SFig{S1}}
\newcommand{\SFigPFvmaxJijcomparison}{\SFig{S2}}

\newcommand{\SFigPFvModAdamvsNAGhJPiaCijab}{\SFig{S3}}

\newcommand{\SFigPFvModAdamvsRPROPLRhJPiaCijab}{\SFig{S4}}
\newcommand{\SFigPFvmaxJijvsrpd}{\SFig{S5}}

\newcommand{\SFigPFvhJcomparison}{\SFig{S6}}
\newcommand{\SFigPFihJcomparison}{\SFig{S7}}

\newcommand{\SFigsPFvPFihJcomparison}{\SFigs{S6 and S7}}

\newcommand{\SFigsPFvPFiKL}{\SFigs{S8}}

\newcommand{\SFigsPFvPFiPiaCijab}{\SFigs{S9 and S10 }}

\newcommand{\SFigPFvMCPiaCijab}{\SFig{S11}}
\newcommand{\SFigPFiMCPiaCijab}{\SFig{S12}}

\newcommand{\SFigsPFvMCPFiMCPiaCijab}{\SFigs{S11 and S12}}

}{

\newcommand{\SFigPFvModAdamvsAdamhJPiaCijab}{\SFig{S1}}
\newcommand{\SFigPFvmaxJijcomparison}{\SFig{S2}}

\newcommand{\SFigPFvModAdamvsNAGhJPiaCijab}{\SFig{S3}}

\newcommand{\SFigPFvModAdamvsRPROPLRhJPiaCijab}{\SFig{S4}}
\newcommand{\SFigPFvmaxJijvsrpd}{\SFig{S5}}

\newcommand{\SFigPFvhJcomparison}{\SFig{S6}}
\newcommand{\SFigPFihJcomparison}{\SFig{S7}}

\newcommand{\SFigsPFvPFihJcomparison}{\SFigs{S6 and S7}}

\newcommand{\SFigsPFvPFiKL}{\SFigs{S8}}

\newcommand{\SFigsPFvPFiPiaCijab}{\SFigs{S9 and S10 }}

\newcommand{\SFigPFvMCPiaCijab}{\SFig{S11}}
\newcommand{\SFigPFiMCPiaCijab}{\SFig{S12}}

\newcommand{\SFigsPFvMCPFiMCPiaCijab}{\SFigs{S11 and S12}}

}% TCBB

}% withSup

\withSup {

\newcommand{\SecPotts}{methods \ref{section: Potts}}
\newcommand{\SecSampleAve}{methods \ref{section: sample_ave}}
\newcommand{\SecEnsembleAve}{methods \ref{section: ensemble_ave}}
\newcommand{\SecSampleAndEnsembleAve}{methods \ref{section: sample_ave} and \ref{section: ensemble_ave}}
\newcommand{\SecEvolution}{methods \ref{section: evolution}}
\newcommand{\SecTsTgTm}{methods \ref{section: Ts_Tg_Tm}}
\newcommand{\SecEvolutionAndTsTgTm}{methods \ref{section: evolution} and \ref{section: Ts_Tg_Tm}}
\newcommand{\SecBML}{methods \ref{section: BML}}
\newcommand{\SecReg}{methods \ref{section: Regularization}}
\newcommand{\SecParamUpdates}{methods \ref{section: Parameter_updates}}
\newcommand{\SecIterations}{methods \ref{section: Iterations}}
\newcommand{\SecGauge}{methods \ref{sec: Ising_gauge_for_comparison}}

\newcommand{\EQelasticNet}{\ref{\EQ: elastic_net}}
\newcommand{\EQgroupL}{\ref{\EQ: group_L1}}
\newcommand{\EQdefSampleAvepsi}{\ref{\EQ: def_sample_ave_of_psi}}
\newcommand{\EQsampleAvepsi}{\ref{\EQ: sample_ave_of_psi}}
\newcommand{\EQensAvepsi}{\ref{\EQ: ensemble_ave_of_psi}}

\newcommand{\EQIsingGauge}{\ref{\EQ: Ising gauge}}

\newcommand{\EQModAdam}{\ref{\EQ: ModAdam}}
\newcommand{\EQgradh}{\ref{\EQ: gradients-h}}
\newcommand{\EQgradJ}{\ref{\EQ: gradients-J}}

\newcommand{\EQKLA}{\ref{\EQ: KL1}}
\newcommand{\EQKLB}{\ref{\EQ: KL2}}

\newcommand{\EQdefGammaij}{\ref{\EQ: def_gamma_ij}}

\newcommand{\EQTs}{\ref{\EQ: equilibrium_distr_of_seq_of_dG}}

\newcommand{\EQequilDistrOfSeqOfdG}{\ref{\EQ: equilibrium_distr_of_seq_of_dG}}

}{

}% withSup

\TCBB{

\withSup {
}{
\newcommand{\SecPotts}{methods S.1.1}

\newcommand{\SecSampleAve}{methods S.1.2}
\newcommand{\SecEnsembleAve}{methods S.1.3}
\newcommand{\SecSampleAndEnsembleAve}{methods S.1.2 and S.1.3}
\newcommand{\SecEvolution}{methods S.1.4}
\newcommand{\SecTsTgTm}{methods S.1.5}
\newcommand{\SecEvolutionAndTsTgTm}{methods S.1.4 and S.1.5}
\newcommand{\SecBML}{methods S.1.6}
\newcommand{\SecReg}{methods S.1.7}
\newcommand{\SecParamUpdates}{methods S.1.8}
\newcommand{\SecIterations}{methods S.1.8.3}
\newcommand{\SecGauge}{methods S.1.9}

\newcommand{\EQelasticNet}{S46}
\newcommand{\EQgroupL}{S54}
\newcommand{\EQdefSampleAvepsi}{S5}
\newcommand{\EQsampleAvepsi}{S6}

\newcommand{\EQensAvepsi}{S9}

\newcommand{\EQIsingGauge}{S72}

\newcommand{\EQModAdam}{S60}
\newcommand{\EQgradh}{S40}
\newcommand{\EQgradJ}{S41}

\newcommand{\EQKLA}{S71}
\newcommand{\EQKLB}{S70}

\newcommand{\EQdefGammaij}{S57}

\newcommand{\EQTs}{S12}

\newcommand{\EQequilDistrOfSeqOfdG}{S12}

}% withSup

}{

\withSup {
}{
}% withSup

}% TCBB
% End of macro-EQ.tex
% End of macro-sup.NAG.tex
\else
\fi

\renewcommand{\TEXT}[1]{#1}
\renewcommand{\SUPPLEMENT}[1]{}
\renewcommand{\SkipSupplToMerge}[1]{}

\TEXT{

\renewcommand{\TEXT}[1]{#1}
\renewcommand{\SUPPLEMENT}[1]{}

\TCBB{
\renewcommand{\EQ}{seq}
\renewcommand{\EQ}{eq}
}{
}% TCBB

\TCBB{
}{
\SECTION{Introduction}
}% TCBB

The inverse Potts problem to estimate
evolutionary fields ($\{h_i\}$) at each site $i$ and couplings ($\{J_{ij}\}$) between each site pair ($i$, $j$)
in the maximum entropy model for the distribution of homologous proteins in sequence space,
$P(\VECS{\sigma}) \propto \exp(- \psi(\VECS{\sigma}))$, where
$\psi \equiv - (\sum_i h_i(\sigma_i) + \sum_{j > i} J_{ij}(\sigma_i, \sigma_j))$,
sequence $\VECS{\sigma} \equiv (\ldots, \sigma_i, \ldots)$,
and $\sigma_i \in \{\text{amino acids, deletion}\}$,
from single-site and pairwise amino acid frequencies
in the multiple sequence alignment (MSA) of homologous proteins
is one of useful methods
in the studies of protein structure and evolution.

For analyses requiring computationally faster methods,
approximate methods such as the mean field approximation\CITE{LGJ:02,LGJ:12,MPLBMSZOHW:11,MCSHPZS:11} and
pseudo-likelihood maximization methods\CITE{ELLWA:13,EHA:14} have been employed.
The performance of contact residue pair prediction by the mean-field or pseudo-likelihood maximization method
is sufficiently good\CITE{CASP11:16}, but it has been reported\CITE{BLCC:16,CFFMW:17,FBW:18} that these methods
can reproduce the structure of a interaction network between amino acid sites in protein structures 
but typically not well reproduce pairwise amino acid frequencies.
A method that can 
better reproduce sequence statistics
\CITE{BLCC:16,CFFMW:17,FBW:18}
is the Boltzmann machine (BM) method\CITE{HS:83,AHS:85,H:07,WWSHH:09,M:17,M:19,M:20}.
In the Boltzmann machine learning (BML),
pairwise marginal distributions are estimated by 
the Markov chain Monte Carlo (MCMC) samplings with the Metropolis-Hastings algorithm\CITE{MRRTT:53,H:70}
or Gibbs sampler\CITE{GG:84},
and the fields and couplings are iteratively estimated 
by maximizing log-likelihood, equivalently by minimizing cross entropy.
Estimating the ensemble averages by simple MCMC is computationally very intensive.
Here, we try to reduce computational time by using persistent MCMC, which
was proposed for Restricted Boltzmann Machines (RBM)
\CITE{H:02, HOT:06, BD:07,T:08}. 
Also, parallel MCMC is employed.
These devices significantly reduce computational time to estimate 
the ensemble averages in each step of the BM learning.
The native protein homologous sequences rather than random sequences are employed as the
initial sequences for the parallel MCMC 
in order not to miss the sequence space around the native sequences.
Because the size of multiple sequence alignment (MSA)(full-batch) is not small, 
stochastic gradient descent methods with mini-batches of about 100 sequences
is employed to reduce computational time for each learning step.

The second problem in
the inverse Potts problem
for estimating evolutionary fields and couplings in protein homologous sequences
is difficulties in adjusting
the values of regularization parameters; there are at least two hyperparameters
to represent
each contribution on the objective function (cross entropy or log-likelihood) of 
the regularization terms for the evolutionary fields
and couplings.
In the case of contact residue pair prediction,
its precision is employed to adjust
the regularization parameters.
However, it
is not sensitive to the regularization parameters 
and not appropriate for estimating evolutionary fields and couplings.
In order to adjust these regularization parameters,
we need to have other measures to determine
which set of values for the regularization parameters yields a more appropriate set of
the fields and couplings.
Here,
these regularization parameters are adjusted for the fields and couplings
to satisfy a specific condition that is appropriate for protein conformations.
This condition is that
the average, $\overline{\psi_N(\VEC{\sigma}_N)}$, of the total native interactions $\psi_N(\VEC{\sigma}_N)$ 
over homologous sequences $\VEC{\sigma}_N$
should be
equal to it's ensemble average, $\langle \psi_N(\VEC{\sigma}) \rangle_{\VEC{\sigma}}$, over the Boltzmann distribution
with the assumption of the Gaussian distribution for interaction density;
$\overline{\psi_N(\VEC{\sigma}_N)} \simeq \langle \psi_N(\VEC{\sigma}) \rangle_{\VEC{\sigma}} = \bar{\psi}_N - \delta\psi_N^2$,
where $\bar{\psi}_N$ and $\delta\psi_N^2$ are the mean and variance of
the Gaussian probability density for the native interactions, 
which are estimated to be the mean and the variance of $\psi_N(\VEC{\sigma})$ over their random sequences.
This condition restricts the appropriate region of the two regularization parameters,
but is not sufficient to select a single point 
in the 2D space of these parameters. 
To do so,  one additional condition is used; 
we select
a set of the regularization parameters yielding the lowest total interactions 
for the native proteins; to take account of the gauge invariance
of interactions, the Ising gauge\CITE{M:19,M:20} is employed.

As examples, we try to estimate evolutionary fields and couplings 
for eight protein families and show the learning profile of each Boltzmann machine, 
which consists of the learning rate schedule, and the changes of the average Kullback-Leibler divergence
of pairwise marginal distributions over all residue pairs,
the mean $\overline{\psi(\VEC{\sigma}_N)}$ of the total interactions 
over the native homologous proteins,
and the ensemble average $\langle \psi(\VEC{\sigma}) \rangle_{\sigma}$.

% End of intro.tex

\SECTION{Methods}

\SUBSECTION{Protein multiple sequence alignments used}
\label{section: MSA}

In order to device a method to optimize regularization parameters in
a Boltzmann machine for estimating 
evolutionary fields and couplings from a protein 
multiple sequence alignment, protein families listed in 
\Table{\ref{\TBL: MSAs}} are employed.

\TableInText{
\ifdefined\ThreeDigitsInTable
\else
\TCBB{
\newcommand{\ThreeDigitsInTable}[2]{#1}
}{
\newcommand{\ThreeDigitsInTable}[2]{#2}
}% TCBB
\fi

\begin{table}[!h]
\begin{threeparttable}[b]
\caption{
\TCBB{
\TEXTBF{
Protein Families Employed.
}
}{
\TEXTBF{
Protein families employed.
}
}% TCBB
}
\label{tbl: MSAs}
\vspace*{1em}
\begin{tabular}{lrlrrrll}
\hline
\\
Pfam ID
	 & $N$ $^{a}$ & $\overline{\text{SeqDiff}}$ $^{b}$ & $N_{\text{eff}}$ $^{c}$ & $N_{\text{rep}}$ $^d$                 & $L$ $^e$	
	& UniProt ID	& PDB ID
	\\
\hline
PF00018			& 17574	& 0.7265  	& 6341.8  	& 7007	& 48		
	& SRC\_HUMAN:90-137	& 1FMK-A:87-134
	\\
PF00127			&  4152	& 0.6808  	& 2503.4  	& 2629	& 128		
	& AZUR\_PSEAE:21-148	& 5AZU-A:1-128
	\\
PF00153	 		& 54582 & 0.7536  	& 19473.9 	& 22061	& 97	
	& UCP2\_MOUSE:112-208	& 2LCK-A:112-208 
	\\
PF00290			&  7687 & 0.6531  	& 3484.1  	& 3762	& 259		
	& TRPA\_ECOLI:8-266 & 1WQ5-A:8-266
	\\
PF00565			& 17571 & 0.7640  	& 9019.9  	& 9451	& 109		
	& (NUC\_STAAU) P00644:115-223 & 1SNC-A:33-141
	\\
PF00595 		& 13814 & 0.7547  	& 4748.8  	& 5149	& 81	
	& PTN13\_MOUSE\_1357-1439:1358-1438	& 1GM1-A:16-96
	\\
PF00887			&  4239	& 0.6916  	& 1806.5  	& 1965	& 80		
	& ACBP\_BOVIN:3-82	& 2ABD-A:2-81
	\\
PF00959			&  2445 & 0.6945  	& 1515.0 	& 1564	& 125		
	& ENLYS\_BPT4:24-148	& 2LZM-A:24-148
	\\
\hline
\end{tabular}
% End of Tables/msas.tex
\begin{tablenotes}
\item [$^a$] Identical sequences are removed.

\item [$^b$] The weighted average of pairwise sequence differences;  
the sequence weight $w_{\VECS{\sigma}_N}$ for a natural sequence $\VECS{\sigma}_N$ is
equal to the inverse of the number of sequences
that are not more than 20\% different from a given sequence.

\item [$^c$] The effective number of sequences, $N_{\text{eff}} \equiv \sum_{\VECS{\sigma}_N} w_{\VECS{\sigma}_N}$,
where the sample weight $w_{\VECS{\sigma}_N}$ for a natural sequence $\VECS{\sigma}_N$ is
equal to the inverse of the number of sequences
that are not more than 20\% different from a given sequence.

\item [$^d$] The number of representative sequences, which are more than 20\% different from any other.

\item [$^e$] The number of residues in a sequence.
% End of Tables/msas_footnotes_3p.tex
\end{tablenotes}
\end{threeparttable}
\end{table}
% End of table_msas.tex
}% TableInText
% End of MSA.tex

% \input{inversePotts.tex}

\SUBSECTION{The inverse Potts model for protein homologous sequences}
\label{section: Potts}

Let us consider probability distributions $P(\VECS{\sigma})$ of
amino acid sequences 
$\VECS{\sigma} (\equiv (\sigma_1, \ldots, \sigma_L) )$,
which satisfy the following constraints that 
single-site and two-site marginal probabilities must be equal to
a given frequency $P_i(a_k)$
of amino acid $a_k$ at each site $i$ and a given frequency $P_{ij}(a_k,a_l)$ of
amino acid pair $(a_k,a_l)$ for site pair $(i,j)$, respectively;
where $\sigma_i, a_k \in \{ \text{amino acids, deletion} \}$.
The sequence distribution 
$P(\VECS{\sigma})$	
with the maximum entropy 
can be represented as
\begin{eqnarray}
   P(\VECS{\sigma})
	= \frac{1}{Z_{\VECS{\sigma}}} e^{- \psi(\VECS{\sigma}) }
	\label{\EQ: max_entropy_distr}
  \hspace*{1em} , \hspace*{1em} Z_{\VECS{\sigma}} =
        \sum_{\VECS{\sigma}} e^{-\psi(\VECS{\sigma}) }
        \label{\EQ: Z_for_Potts}
	\\
  \psi(\VECS{\sigma})
        = - \, [ \,
		\sum_i \{ \, h_i(\sigma_i) + \sum_{j(>i)} J_{ij}(\sigma_i, \sigma_j) \, \}
		\, ]
        \label{\EQ: H_for_Potts}
        \label{\EQ: psi_for_Potts}
\end{eqnarray}
where Lagrange multipliers $h_i(a_k)$ and $J_{ij}(a_k,a_l)$ are 
interaction potentials called fields and couplings,
and 
$\psi(\VECS{\sigma})$ 
may be called evolutionary energy.

The fields $h_i(a_k)$ and couplings $J_{ij}(a_k,a_l)$ provide 
useful information to understand protein evolution\CITE{M:17,M:19,M:20}
and also to predict residue-residue contacts in protein structures 
on the basis of coevolutional residue substitutions \CITE{LGJ:02,MPLBMSZOHW:11,MCSHPZS:11,M:13,M:17b,M:18}.
Estimating $h_i(a_k)$ and $J_{ij}(a_k,a_l)$ 
from given
single-site $P_{i}(a_k)$ and two-site frequencies $P_{ij}(a_k,a_l)$,
which are evaluated from a multiple sequence alignment,
have been attempted as the inverse Potts problem
by the mean field approximation\CITE{LGJ:02,MPLBMSZOHW:11,MCSHPZS:11}, 
by the Gaussian approximation\CITE{BZFPZWP:14},
by maximizing a pseudo-likelihood\CITE{ELLWA:13,EHA:14,BKCLL:11,KOB:13},
by minimizing a cross entropy in the adaptive cluster expansion\CITE{BLCC:16},
by the Boltzmann machine (BM) method\CITE{WWSHH:09,FBW:18},
and
by the restricted Boltzmann machine (RBM) method\CITE{TCM:19,SW:19}.
Here we argue BM learning to estimate the interactions, $\{h_i(a_k)\}$ and $\{J_{ij}(a_k,a_l)\}$, in homologous protein sequences.

% End of inversePotts.tex
% \input{BMLearning.tex}

\newcommand{\norm}[1]{\left\lVert#1\right\rVert}
\newcommand{\argmin}{\text{argmin}}
\newcommand{\smalltext}[1]{\text{\script{#1}}}

\SUBSECTION{Boltzmann machine}
\label{section: BML}

The cross entropy with a regularization term, $S$, which corresponds to a negative log-posterior-probability per instance, is minimized.
\begin{eqnarray}
	S &\equiv& \frac{- 1}{\sum_{\tau} 1} \sum_{\tau} \log P(\VECS{\sigma}^{\tau}) + R
	\label{\EQ: cross_entropy}
\end{eqnarray}
where $R$ is a regularization term, and $\tau$ denotes an instance. 
According to \CITE{FBW:18}, instead of $\{h_i\}$ and $\{J_{ij}\}$, 
we use the new parameters $\{\phi_{i}\}$ and $\{\phi_{ij}\}$ for minimization, 
which are Lagrange multipliers
corresponding to $[ \sum_{\VECS{\sigma}}P(\VECS{\sigma})\delta_{\sigma_i a_k} - P_i(a_k) ]$ and
$[ \sum_{\VECS{\sigma}} P(\VECS{\sigma}) \delta_{\sigma_i a_k} \delta_{\sigma_j a_l} 
- P_{ij}(a_k,a_l) 
- \sum_{\VECS{\sigma}} P(\VECS{\sigma}) \delta_{\sigma_i a_k} P_j(a_l)
- P_i(a_k) \sum_{\VECS{\sigma}} P(\VECS{\sigma}) \delta_{\sigma_j a_l}
+ 2 P_i(a_k) P_j(a_l) ] $
in the maximum entropy model, respectively.
The partial derivatives of the cross entropy can be easily calculated:
\begin{eqnarray}
	\frac{\partial S}{\partial \phi_i(a_k)} &=& \sum_{\VECS{\sigma}} P(\VECS{\sigma}) \delta_{\sigma_i a_k}  - P_i(a_k)
					+ \frac{\partial R}{\partial \phi_i(a_k)} 
	\label{\EQ: gradients-h}
		\\
	\frac{\partial S}{\partial \phi_{ij}(a_k, a_l)} &=& 
	[
	\sum_{\VECS{\sigma}} P(\VECS{\sigma}) \delta_{\sigma_i a_k} \delta_{\sigma_j a_l} - P_{ij}(a_k,a_l) 
	\nonumber
	\\
 	& &
	- \sum_{\VECS{\sigma}} P(\VECS{\sigma}) \delta_{\sigma_i a_k} P_j(a_l)
	- P_i(a_k) \sum_{\VECS{\sigma}} P(\VECS{\sigma}) \delta_{\sigma_j a_l}
	\hspace*{1em}
	\nonumber
	\\
	& &
	+ 2 P_i(a_k) P_j(a_l) ]
	]
					+ \frac{\partial R}{\partial \phi_{ij}(a_k, a_l)} 
	\label{\EQ: gradients-J}
\end{eqnarray}
The relationships between $(h_i, J_{ij})$ and $(\phi_i, \phi_{ij})$
are as follows.
\begin{eqnarray}
	h_i(a_k) &=& \phi_i(a_k) - \sum_{j (\neq i) } \sum_l \phi_{ij}(a_k, a_l)) P_j(a_l)
	\\
	J_{ij}(a_k, a_l) &=& \phi_{ij}(a_k, a_l)
\end{eqnarray}

The single-site and two-site frequencies, $P_i(a_k)$ and $P_{ij}(a_k, a_l)$, are evaluated as follows from
homologous sequences, each of which has a sample weight $w_{\VECS{\sigma}_{N}}$,
in a multiple sequence alignment.
\begin{eqnarray}
	P_i(a_k) &=& \sum_{\VECS{\sigma}_N} w_{\VECS{\sigma}_N} \delta_{\sigma_{N i} a_k} / \sum_{\VECS{\sigma}_N} w_{\VECS{\sigma}_N}
		\label{\EQ: Pi_for_BM}
		\\
	P_{ij}(a_k,a_l) &=& \sum_{\VECS{\sigma}_N} w_{\VECS{\sigma}_N} \delta_{\sigma_{N i} a_k} \delta_{\sigma_{N j} a_l} / \sum_{\VECS{\sigma}_N} w_{\VECS{\sigma}_N}
		\label{\EQ: Pij_for_BM}
\end{eqnarray}
where $\VECS{\sigma}_N$ denotes natural sequences.
The sample weight $w_{\VECS{\sigma}_N}$ for each homologous sequence, 
which is used to reduce phylogenetic biases in the set of the homologous sequences,
is defined here to be equal to 
the inverse of the number of sequences that are not more than 20\% different from a given sequence,
including itself.

\SUBSECTION{Regularization}
\label{section: Regularization}

Couplings $\phi_{ij}(a_k, a_l)$ are expected to be 
significant between the residues that are closely located 
in a 3D protein structure and complex.
Thus, they are expected to be sparse, because
the number of residue-residue contacts in a protein 3D structure 
is typically between 2 and 4 per residue depending on a criterion,
in comparison with the number of residue pairs, $L(L-1)/2$, 
where $L$ is a protein length\CITE{MJ:82}.
Here, to take account of the sparsity of the couplings, 
Group $L_1$ is employed to deal with couplings, $\phi_{ij}(a_k, a_l)$, 
between residues $i$ and $j$ as a group.  
For single-site interactions, $\phi_{i}(a_k)$, $L_2$ regularization is employed.
\begin{equation}
R \equiv
	\lambda_1 \sum_i \sum_k \frac{1}{2} \{ \phi_i(a_k)^2 \}
	+
	\lambda_2 \sum_i \sum_{j (> i) } \sqrt{\sum_k \sum_l \{ \phi_{ij}(a_k, a_l)^2 \} }
	\hspace*{1em}
	\label{\EQ: group_L1}
\end{equation}
where $\lambda_1, \lambda_2 \geq 0$.
For minimizing the cross entropy of \Eq{\ref{\EQ: cross_entropy}} including a group L1 regularization term, 
the soft-thresholding function\CITE{Sm:17.540L5} is employed\CITE{M:19}.
% End of BMLearning.tex
% \input{Bayesian_correction.tex}

\SUBSECTION{Bayesian corrections for single-site and pairwise amino acid frequencies in data and in MCMC samples}

The amino acid frequencies $f_i(a_k)$ at each site and pairwise frequencies $f_{ij}(a_k, a_l)$ between each site pair 
in the observed data or in the MCMC samples may be corrected in the Bayesian manner.
\begin{eqnarray}
    \hat{f}_{i}(a_k) &=& ( {f}_i(a_k) N_{\text{eff}}  + N_{\text{pseudo}} / |\{a_k\}| \,) / (N_{\text{eff}} + N_{\text{pseudo}} )
	\label{\EQ: BayesianCorrection1}
	\\
    \hat{f}_{ij}(a_k, a_l) &=& ( {f}_{ij}(a_k, a_l) N_{\text{eff}} + N_{\text{pseudo}} / |\{a_k\}|^2 \,) / (N_{\text{eff}} + N_{\text{pseudo}} ) 
	\label{\EQ: BayesianCorrection2}
\end{eqnarray}
where $N_{\text{eff}}$ is the effective number of sequences in the data
and $N_{\text{pseudo}}$ is a pseudo-count that may take a different value depending on each case.

% End of Bayesian_correction.tex
% \input{MCMC.tex}

\SUBSECTION{Markov chain Monte Carlo method}

Given $\sigma_{-i} \equiv \{ \sigma_{j (\neq i)} \}$, the conditional probability $P(\sigma_i = a_k | \sigma_{-i} )$
can be represented as
\begin{eqnarray}
  P(\sigma_i = a_k | \sigma_{-i} ) &=&
	\frac{\exp (h_i(a_k) + \sum_{j \neq i} J_{ij}(a_k, \sigma_j) ) } {\sum_{l} \exp (h_i(a_l) + \sum_{j \neq i} J_{ij}(a_l, \sigma_j) ) }
\end{eqnarray}
Thus, the Gibbs sampler\CITE{GG:84} can be employed for the MCMC method. However, the number of amino acid types 
including deletion, $| \{ a_k \} |$, is equal to 21 and evaluating
the conditional probabilities for all amino acid types at each site $i$ requires 
non-negligible computational time.
The alternative algorithm is the use of the Metropolis-Hastings algorithm\CITE{H:70}.
The acceptance probability $\alpha_i$ for $a_k$ at the $i$-th site is
\begin{eqnarray}
	\alpha_i(\sigma_i, a_k) &=& \min \{ \frac{ P(\sigma_i = a_k | \sigma_{-i} ) q(a_k, \sigma_i | \sigma_{-i} )}
			{  P(\sigma_i  | \sigma_{-i} ) q(\sigma_i, a_k | \sigma_{-i} ) } , 1 \}
\end{eqnarray}
where $q(\sigma_i, a_k  | \sigma_{-i} )$ is the proposed probability of $a_k$ for $\sigma_i$.
For simplicity, the amino acid frequency,
which is 
calculated from native sequences with sample weights by \Eq{\ref{\EQ: Pi_for_BM}}
and corrected with a pseudo-count $N_{\text{pseudo}}=10$, 
at each site in data is used here as the proposed probability; $q(\sigma_i, a_k  | \sigma_{-i} ) =  \hat{P}_{i}(a_k)$ ; 
 see \Eq{\ref{\EQ: BayesianCorrection1}} for the Bayesian correction of the the proposed probabilities.

% End of MCMC.tex
% \input{persistentMCMC.tex}

\SUBSECTION{Parallel, persistent Markov chain Monte Carlo method to estimate stochastic gradients of the cross entropy}

To minimize the cross entropy of a Boltzmann machine (BM), 
its partial gradients, \Eqs{\Ref{\EQ: gradients-h} and \Ref{\EQ: gradients-J}}, must be evaluated.
However, there are intractable terms in their equations which require  
the model's expectations of amino acid pairwise frequencies.

The Contrasted Divergence approximation (CD) was introduced 
as a fast learning approximation to 
minimize
the cross entropy of Restricted Boltzmann Machines (RBM);
more specifically for the CD-1\CITE{H:02, HOT:06, BD:07}, but
it was also suggested that the CD-$k$, CD between Kullback-Leibler divergences, $P_t \parallel P_{\infty}$ with $t=0$ and $t=k$, in the Markov chain, for greater $k$ was preferred over CD-1;
$P_0$ is the probability distribution of data.
The persistent CD algorithm\CITE{T:08} was then introduced;
instead of running a Markov chain with many steps to estimate the equilibrium ensemble of each model, 
a Markov chain is initialized at the state in which it ended for a previous model. 
In this approximation it is assumed that the model changes only slightly between parameter updates.
Originally this persistent algorithm was introduced for the RBM but it can be
applied to the BM learning.

Here the parallel, the persistent MCMC method shown in \Fig{\ref{fig: persistentMCMC}} 
is employed for the BM learning.
Each Markov chain, $P_t$, asymptotically converges to the model's probability distribution,
$\lim_{t \rightarrow \infty} P_t = P_{\infty}$, but
in practice a Markov chain of finite time must be used.  
It is often attempted to reduce the bias by removing some "burn-in", but
this never completely removes the bias. Asymptotically, as $t \rightarrow \infty$, the bias becomes negligible,
but it is hard to know how large $t$ needs to be\CITE{W:19}.  
Averaging biased estimators of parallel Markov chains may be more biased
than time average in a single MCMC\CITE{W:19}, because 
the expectation of a finite time Markov chain is biased\CITE{GR:14}.
However, the sequence space for proteins is equal to $21^L$ including the amino acid deletion, 
where $L$ is the sequence length.  Parallel Markov chains would walk with less computational time in larger sequence space
than a single Markov chain does.  Here the native homologous sequences aligned are employed as the initial sequences for 
the Boltzmann machine not to miss sequence space around the native sequences; 
the representative sequences, which are more than 20\% different from any other, in each Protein MSA
are employed for the MCMC method and called here as a full-batch.  
After the sequence order of the full-batch is shuffled, it is divided into mini-batches, 
which contains sequences whose number is mostly between 100 and 110 or more than that for the last one in the present work.
Each mini-batch is employed to generate Markov chains in parallel,
and the $k$ steps/residues of MCMC are executed before evaluating model's averages of pairwise amino acid occurrences 
and then to update parameters; $k = 10$ was employed in the present work. 
After the full-batch is processed, and 
then the whole process is iterated again by
shuffling the sequence order of the full-batch.
Here it should be noted that each sequence in the full-batch starts from the state in which it ended for a previous model.

\FigureInText{
\begin{figure*}[!h]
\centerline{
\includegraphics[width=180mm,angle=0]{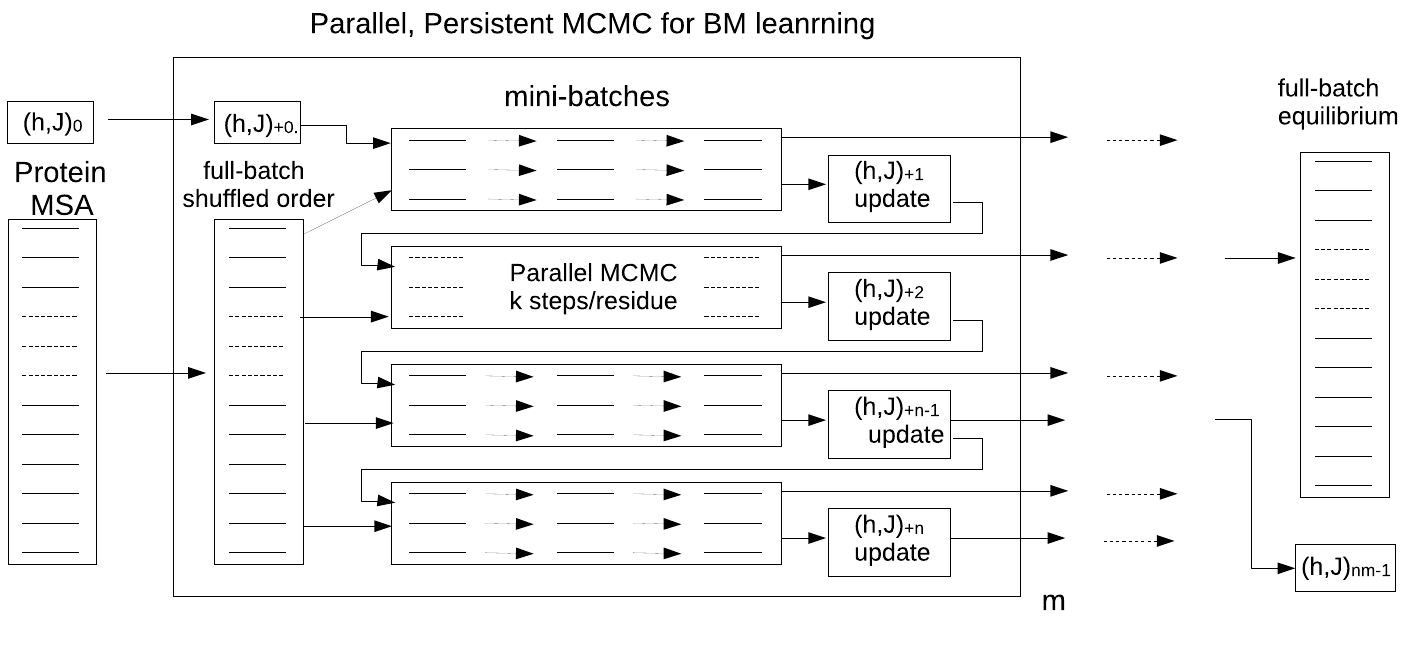}
}
\caption
{
\TEXTBF{A schematic representation of parallel, persistent Markov chain Monte Carlo method for the Boltzmann machine learning}
\label{fig: persistentMCMC}
}
\end{figure*}

% End of fig_persistentMCMC.tex
\clearpage
}% FigureInText

% End of persistentMCMC.tex

% \input{Strategy_to_get_best.tex}

\SUBSECTION{Stochastic gradient descent methods (SGD)}

The adaptive learning rate method, Adam\CITE{KB:14}, has been employed here as well as 
the modified version of Adam, which is called ModAdam\CITE{M:19,M:20};
the hyperparameters in Adam and ModAdam are taken to be
$\epsilon = 1.0^{-6}$, and 
$\beta_1 = 0.9$ and $\beta_2 = 0.999$ according to Adam\CITE{KB:14}.

\SUBSECTION{Initialization of the interactions, $\{ h_i(a_k) \}$ and $\{ J_{ij}(a_k, a_l) \}$ }

    The interactions, $\{ h_i(a_k) \}$ and $\{ J_{ij}(a_k, a_l) \}$, are initialized as follows.
\begin{eqnarray}
    h_{i}(a_k) &\equiv& \log \hat{P}_{i}(a_k) - \frac{1}{|\{a_k\}|} \sum_k \log \hat{P}_{i}(a_k)
	\\
    J_{ij}(a_k, a_l) &\equiv& \mathcal{N}(0, \sigma^2) \hspace*{2em} \text {with } \sigma = 10^{-3}
\end{eqnarray}
where initial values for the couplings $J_{ij}(a_k, a_l)$ are drawn from the normal distribution 
$\mathcal{N}(\mu, \sigma^2)$ with the mean $\mu$ and the standard deviation $\sigma$.
The amino acid frequencies $\hat{P}_i(a_k)$ at each site for the initialization of $h_i(a_k)$
are estimated with a pseudo-count, $N_{\text{pseudo}} = 10$; see \Eq{\ref{\EQ: BayesianCorrection1}}.

\SUBSECTION{Estimators of single-site and pairwise marginal distributions from MCMC samples}

The single-site and pairwise marginal distributions in \Eqs{\ref{\EQ: gradients-h} and \ref{\EQ: gradients-J}} are estimated from MCMC samples as follows.
\begin{eqnarray}
  \sum_{\VECS{\sigma}} \hat{P}(\VECS{\sigma}) \delta_{\sigma_i a_k} &=&
        \frac{1}{| \mathcal{S} |}
        \sum_{\VECS{\sigma} \in \mathcal{S}} \delta_{\sigma_i a_k}
        \\
  \sum_{\VECS{\sigma}} \hat{P}(\VECS{\sigma}) \delta_{\sigma_i a_k} \delta_{\sigma_j a_l} &=&
        \frac{1}{| \mathcal{S} |}
        \sum_{\VECS{\sigma} \in \mathcal{S}} \delta_{\sigma_i a_k} \delta_{\sigma_j a_l}
\end{eqnarray}
where $\mathcal{S}$ is the set of MCMC samples, whose number is equal to the size of the mini-batch.
It should be noted here that these estimators are not corrected with a pseudo-count; $N_{\text{pseudo}} = 0$.

\SUBSECTION{Learning schedule}

The learning rate schedule consists of three stages for each SGD. The first stage is a warming-up stage in which the learning rate
linearly increase from 0 to the maximum learning rate $\kappa_{\text{max}}$.  The second stage is the learning stage with
the maximum learning rate. The last stage is the decay stage of the learning rate from $\kappa_{\text{max}}$ to 0.  
Thus, the learning rate changes with $t$ in the following manner; $t$ is the count of parameter updates.
\begin{eqnarray}
	\kappa_0(t) &=&   \left\{ \begin{array}{ll}
	 	(\kappa_{\text{max}} / t_{\text{warmup}} ) \, t	 & \text{ for  } 0 \leq t \leq t_{\text{warmup}} \\
	 	\kappa_{\text{max}}	 & \text{ for  } t_{\text{warmup}} \leq t \leq t_{\text{learning}}\\
	 	\kappa_{\text{max}} ( 1 + a(t - t_{\text{learning}}) )^b & \text{ for  } t_{\text{learning}} \leq t
				\end{array}
			\right.
\end{eqnarray}
where $a > 0$ and $b < 0$; $t_{\text{warmup}} = 100$, $t_{\text{learning}} = 1800$, 
$a = 0.01$, and $b = -0.5$ are employed in the present work.

After the learning for $t_{\text{learning}}$ steps by the ModAdam,
the learning by the Adam follows with the learning rate decay.
In stochastic gradient descent methods with a constant stepsize, 
the expected distance to solution does not converge to 0, and
the asymptotic value decreases as the stepsize decreases\CITE{Sm:17.540L6}.
Therefore the stage of learning rate decay is added in the last SGD, although
the effective learning rate in the Adam and ModAdam is not constant.

\SUBSECTION{Monitoring the learning process}
\label{section: monitoring}

It is useful to visualize how the objective function, cross entropy, changes 
in the minimization process. 
However, the objective function to be minimized can hardly be evaluated for the Boltzmann machine, 
although its partial-derivatives can be easily calculated and then it can be minimized. 
Therefore, we 
alternatively monitor the average, $D_{2}^{KL}$, of Kullback-Leibler divergences for pairwise marginal distributions 
over all residue pairs, which is 
a measure of the similarity of pairwise marginal distributions between the reference and model.
\begin{equation}
D_{2}^{KL} 
	\equiv
	 \frac{2}{L(L-1)} \sum_{i} \sum_{j > i} 
		\sum_{k} \sum_{l} P_{ij}(a_k, a_l) \log \frac{ P_{ij}(a_k, a_l) }
			{ \sum_{\VECS{\sigma} \in \mathcal{S}} \delta_{\sigma_i a_k} \delta_{\sigma_j a_l} / | \mathcal{S} | }
	\label{\EQ: KL2}
\end{equation}
\begin{equation}
	D_{1}^{KL} \equiv \frac{1}{L} \sum_{i}
		\sum_{k} P_{i}(a_k) \log \frac{ P_{i}(a_k) }
	 	{ \sum_{\VECS{\sigma} \in \mathcal{S}} \delta_{\sigma_i a_k} / | \mathcal{S} | }
	\hspace*{5em}
	\label{\EQ: KL1}
\end{equation}
In the case of $P_{i}(a_k) = 0$ or $P_{ij}(a_k, a_l) = 0$, their terms are taken to be equal to $0$.
The single-site and pairwise marginal distributions in the MCMC samples $\mathcal{S}$ are corrected with 
a pseudo-count, $N_{\text{pseudo}} = 1$, for a small number of samples.

Although a mini-batch can be employed 
for the evaluation of the single-site and pairwise frequencies,
multiple mini-batches in the preceding steps, 
which correspond to about 10000 samples,
are employed
in order to reduce the statistical errors due to the small size of the sample set $\mathcal{S}$
and also the fluctuations of $D_{1}^{KL}$ and $D_{2}^{KL}$. 

The learning has been stopped when $\min D_{2}^{KL}$  
does not improve during a certain number, 200, of iterations in the decay stage after $t_{\text{learning}}$ steps.

\SUBSECTION{Comparison of interactions, $\{h_i(a_k)\}$, $\{J_{ij}(a_k, a_l)\}$ and $\psi(\VEC{\sigma})$, between various models}

\label{sec: Ising_gauge_for_comparison}

The $P(\VEC{\sigma})$ of \Eq{\ref{\EQ: max_entropy_distr}} is invariant
under a certain transformation of the fields and couplings,
$J_{ij}(a_k,a_l) \rightarrow J_{ij}(a_k,a_l) - J^1_{ij}(a_k,\ ) - J^1_{ij}(\ ,a_l) + J^0_{ij}$,
$h_i(a_k) \rightarrow h_i(a_k) - h^0_i + \sum_{j\neq i} (J^1_{ij}(a_k,\ ) - J^0_{ij})$ for any
$J^1_{ij}(a_k,\ ) = J^1_{ji}(\ ,a_k)$, $J^0_{ij} = J^0_{ji}$ and $h^0_i$.
Therefore, in order to compare $\{ h_i(a_k) \}$, $\{ J_{ij}(a_k, a_l) \}$ and $\psi(\VEC{\sigma})$ between various models, 
they must be represented at least in a certain gauge. 
Here the following gauge, which we call the Ising gauge\CITE{M:19,M:20}, is employed.
\begin{eqnarray}
        h_i(\cdot) = 0 \:,\hspace{1em} J_{ij}(a_k, \cdot) = 0 \:,\hspace{1em} J_{ij}(\cdot, a_l) = 0
        \label{\EQ: Ising gauge}
\end{eqnarray}
where ``$\cdot$'' denotes the reference state, which is the average over all states 
for the Ising gauge; $h_i(\cdot) \equiv \sum_k h_i(a_k) / (\sum_k 1)$. 
Any gauge can be transformed to this gauge by the following transformation.
\begin{eqnarray}
J^{\script{I}}_{ij}(a_k,a_l) &\equiv& J_{ij}(a_k,a_l) - J_{ij}(a_k,\cdot)
			- J_{ij}(\cdot,a_l) +  J_{ij}(\cdot,\cdot)
		\hspace*{3em}
		\\
h^{\script{I}}_{i}(a_k) &\equiv& h_{i}(a_k) - h_{i}(\cdot) + 
	\sum_{j \neq i} (J_{ij}(a_k, \cdot) - J_{ij}(\cdot, \cdot) )
\end{eqnarray}

% End of Ising-gauge.tex

\SUBSECTION{How to adjust the regularization parameters $\lambda_1$ and $\lambda_2$ as well as the maximum learning rate $\kappa_{\text{max}}$}

In the case of proteins, the prediction accuracy of contact residue pairs in 3D protein structures is
often used as a verification data.  However, the prediction performance is not sensitive to the values of hyperparameters.
The most requirement for protein sequences is ability for folding into unique 3D structures.
The random energy model (REM)\CITE{SG:93a,SG:93b,RS:94,PGT:97} 
and 
the independent interaction model (IIM)\CITE{PGT:97}
for protein folding 
indicate that the energy density in protein conformational space
and correspondingly $\psi$ density in sequence space can be approximated by the Gaussian distribution,
the mean and variance of which are evaluated as ones for random sequences.
Then,
approximating the $\psi$ density in the sequence space
by the Gaussian distribution, we can get the following formula for a relationship between the sample average and ensemble average.
\begin{eqnarray}
	\overline{\psi_N(\VEC{\sigma}_N)} &\equiv& 
		\frac{\sum_{\VEC{\sigma} \in \{ \VEC{\sigma}_N\} } w_{\VEC{\sigma}} \psi_N(\VEC{\sigma}) } 
		{\sum_{\VEC{\sigma}\in \{ \VEC{\sigma}_N \} } w_{\VEC{\sigma}} }
		\\
		&\simeq& \langle \psi_N(\VEC{\sigma}) \rangle_{\VEC{\sigma}}
		\equiv \sum_{\VEC{\sigma}} \psi_N(\VEC{\sigma}) P_N(\VEC{\sigma})	\\
		&\approx&
                          \int \psi \exp ( - \psi ) \mathcal{N}(\bar{\psi}_N, \delta \psi_N^2) d\psi 
			\ / \int \exp ( - \psi ) \mathcal{N}(\bar{\psi}_N, \delta \psi_N^2) d\psi
		\\
		&=& \bar{\psi}_N - \delta \psi_N^2
                \label{\EQ: Gaussian density}
\end{eqnarray}
where $\VEC{\sigma}_N$ is a native sequence in a protein multiple sequence alignment  
and $w_{\VEC{\sigma}}$ is the sample weight to reduce sampling biases in the sequence space;
$w_{\VEC{\sigma}} = 1$ for representative sequences employed here.
The mean and variance, $\bar{\psi}_N$ and $\delta \psi_N^2$, of $\psi_N(\VEC{\sigma})$ for random sequences
are evaluated with the amino acid composition which is
calculated from native sequences with sample weights by \Eq{\ref{\EQ: Pi_for_BM}}.
In \Eq{\ref{\EQ: Gaussian density}} the subscript $N$ is attached for each of $\bar{\psi}$, $\delta \psi^2$, and $\psi$ 
in order to clarify that they are those for the native interactions.
Thus, the relationship above, which should be satisfied by the $\{h_i(a_k)\}$ and $\{J_{ij}(a_k,a_l))\}$ estimated for protein families,
is employed here to adjust the regularization parameters.

First determine the maximum learning rate $\kappa_{\text{max}}$;
$\kappa_{\text{max}}$ for the ModAdam is almost two times larger than that for the Adam.
Let us assume that the interactions, $\{h_i(a_k)\}$ and $\{J_{ij}(a_k, a_l)\}$, are represented in the Ising gauge.
The value of $\bar{\psi} - \delta \psi^2$ in the learning with $\kappa_{\text{max}}$ 
by the second SGD and in the early stage of learning rate decay should almost monotonically decrease.
In the last stage of learning rate decay, $\bar{\psi} - \delta \psi^2$ and $\overline{\psi(\VEC{\sigma}_N)}$ should converge,
otherwise $\kappa_{\text{max}}$ may be slightly large or small 
depending on whether $\bar{\psi} - \delta \psi^2$ and $\overline{\psi(\VEC{\sigma}_N)}$ are increasing or decreasing.
The typical value for $\kappa_{\text{max}}$ for Adam is from 0.005 to 0.0005 in the present case.

The converged values of $\overline{\psi(\VEC{\sigma}_N)}$ and $\bar{\psi} - \delta \psi^2$
are approximately decreasing functions of the regularization constants, $\lambda_1$ and $\lambda_2$.
Because the regularization required would be much weaker for one-body interactions $\{h_i(a_k)\}$ 
than for pairwise interactions $\{J_{ij}(a_k,a_l)\}$, 
an appropriate value for $\lambda_1$
would be less than or equal to the value of $\lambda_2$. 
Thus, the appropriate set of $\lambda_1$ and $\lambda_2$ may be determined by the following procedure.

\begin{enumerate}

\item A range to be searched for $\lambda_2$ is equal to $[ \lambda_{2:L}, \lambda_{2:H} ]$ , which are defined as
\begin{eqnarray}
\lambda_{2:L}	
	&\equiv& \text{argmin}_{\lambda_2}  ( \, \lambda_2 \, | \, \bar{\psi} - \delta \psi^2 \geq \overline{\psi(\VEC{\sigma}_N)} \text{ with } \lambda_1 = \lambda_2 \, ) \, , \, 
	\\
\lambda_{2:H}	
	&\equiv&
	  \text{argmax}_{\lambda_2:h}  ( \, \lambda_{2:h} \, | \, \bar{\psi} - \delta \psi^2 \leq \overline{\psi(\VEC{\sigma}_N)} \text{ for } \lambda_2 \leq \lambda_{2:h} \text{ with } \lambda_1 = 1.0^{-8} \text{ or } 1.0^{-7} \, ) \, ]
\end{eqnarray}

\item  
	$ 
	  (\lambda_1, \lambda_2) =  \text{argmin}_{\lambda_1, \lambda_2} ( \overline{\psi(\VEC{\sigma}_N)} \, | \, \bar{\psi} - \delta \psi^2 \simeq \overline{\psi(\VEC{\sigma}_N)} \, \text{ and } \lambda_{2} \in [ \lambda_{2:L}, \lambda_{2:H} ] )
	$ ;
See \Fig{\ref{\FIG: tuning}}.

\end{enumerate}

The employed values of $\lambda_1$ and $\lambda_2$ as well as
$\kappa_{\text{max}}$ and $t_{\text{learning}}$ for ModAdam and Adam, 
and decay-steps 
are listed in \Table{\ref{\TBL: Parameter_range}}.

\FigureInText{
\ifdefined\DIR
\renewcommand{\DIR}[1]{Figs_prt/#1}
\else
\newcommand{\DIR}[1]{Figs_prt/#1}
\fi
\ifdefined\PRT
\renewcommand{\PRT}{PF00018}
\else
\newcommand{\PRT}{PF00018}
\fi
\ifdefined\SUBDIR
\renewcommand{\SUBDIR}[1]{\DIR{\PRT/#1}}
\else
\newcommand{\SUBDIR}[1]{\DIR{\PRT/#1}}
\fi
\begin{figure*}[!h]
\centerline{
\includegraphics[width=60mm,angle=0]{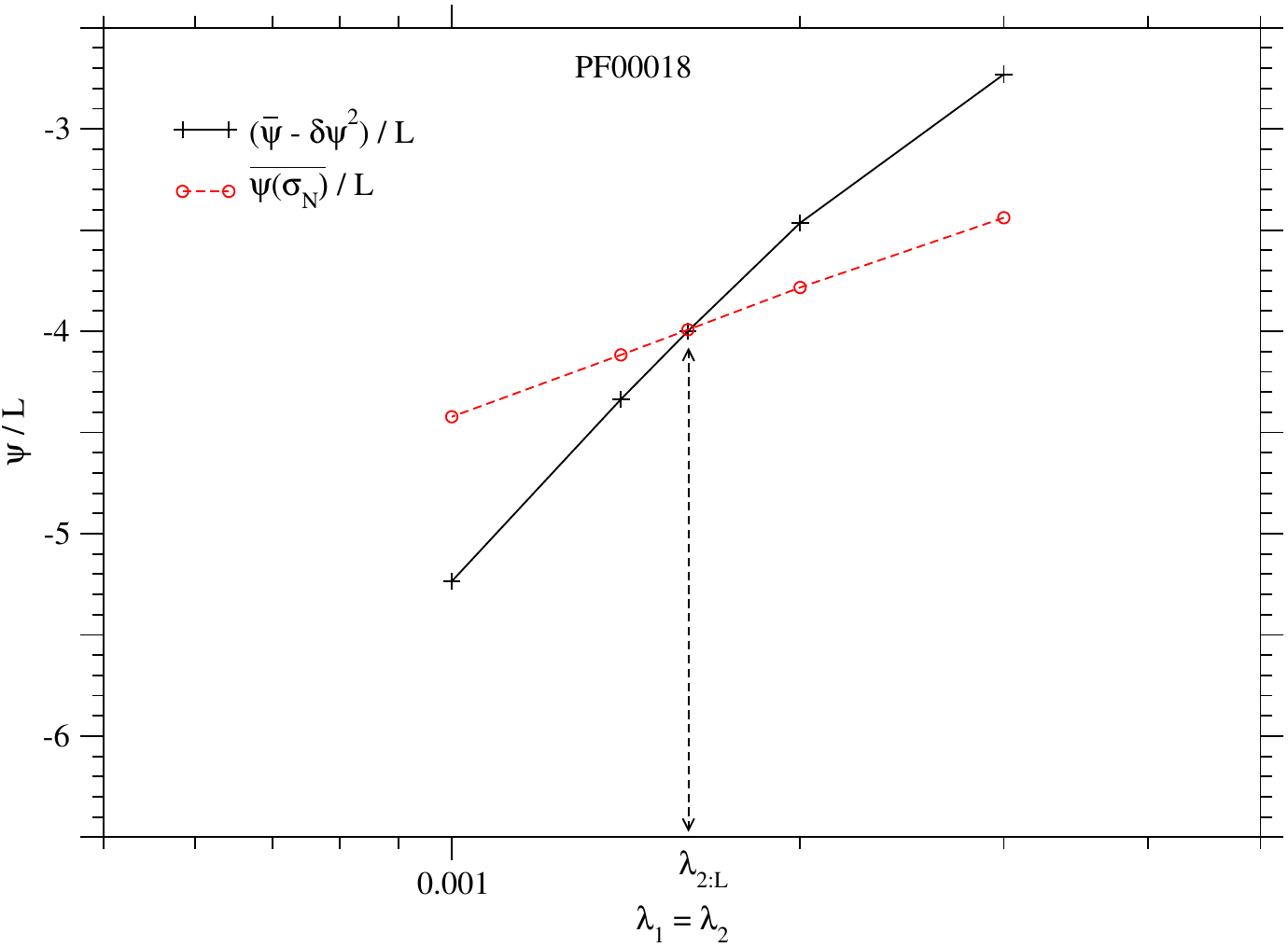}
\includegraphics[width=61mm,angle=0]{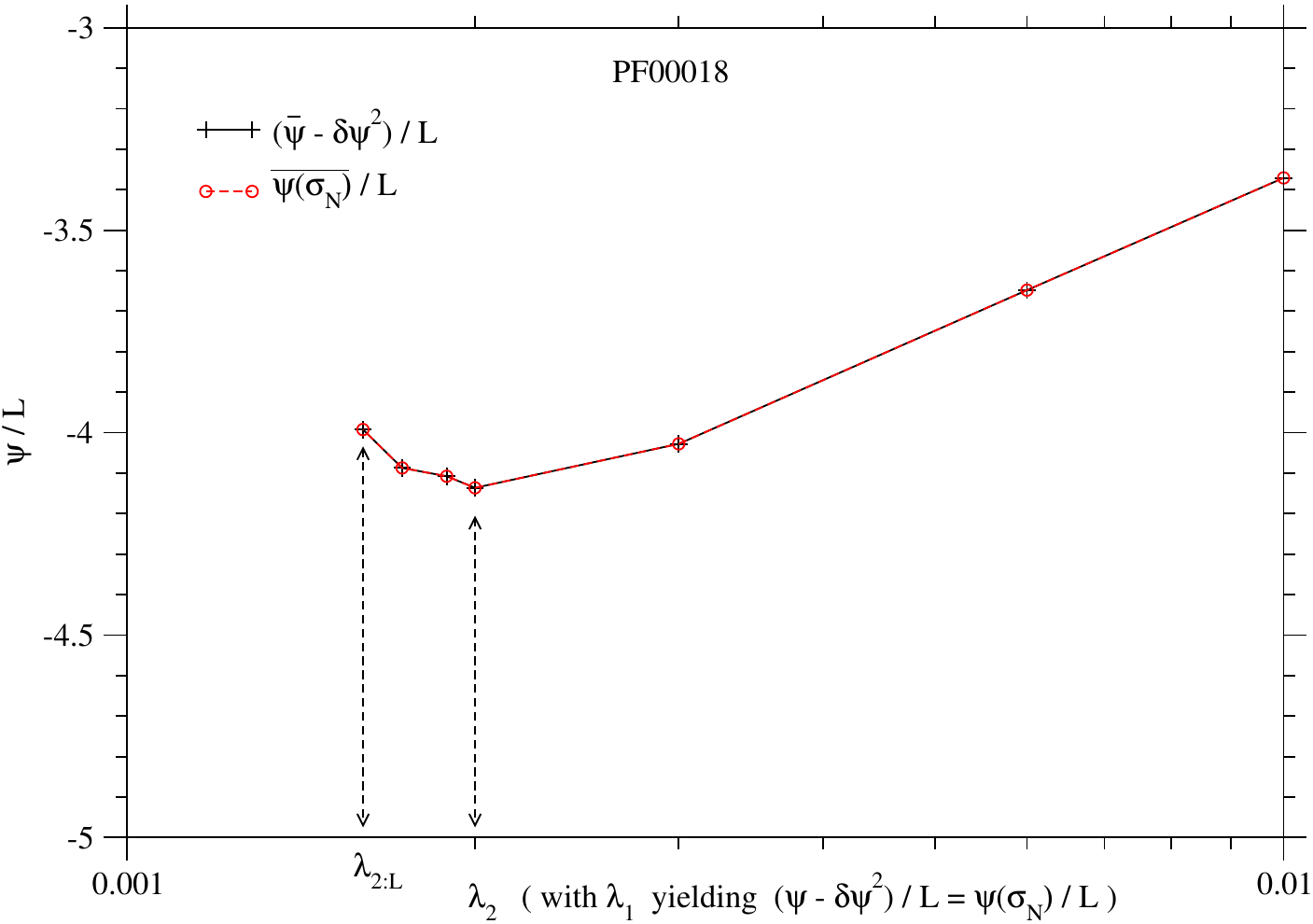}
\includegraphics[width=60mm,angle=0]{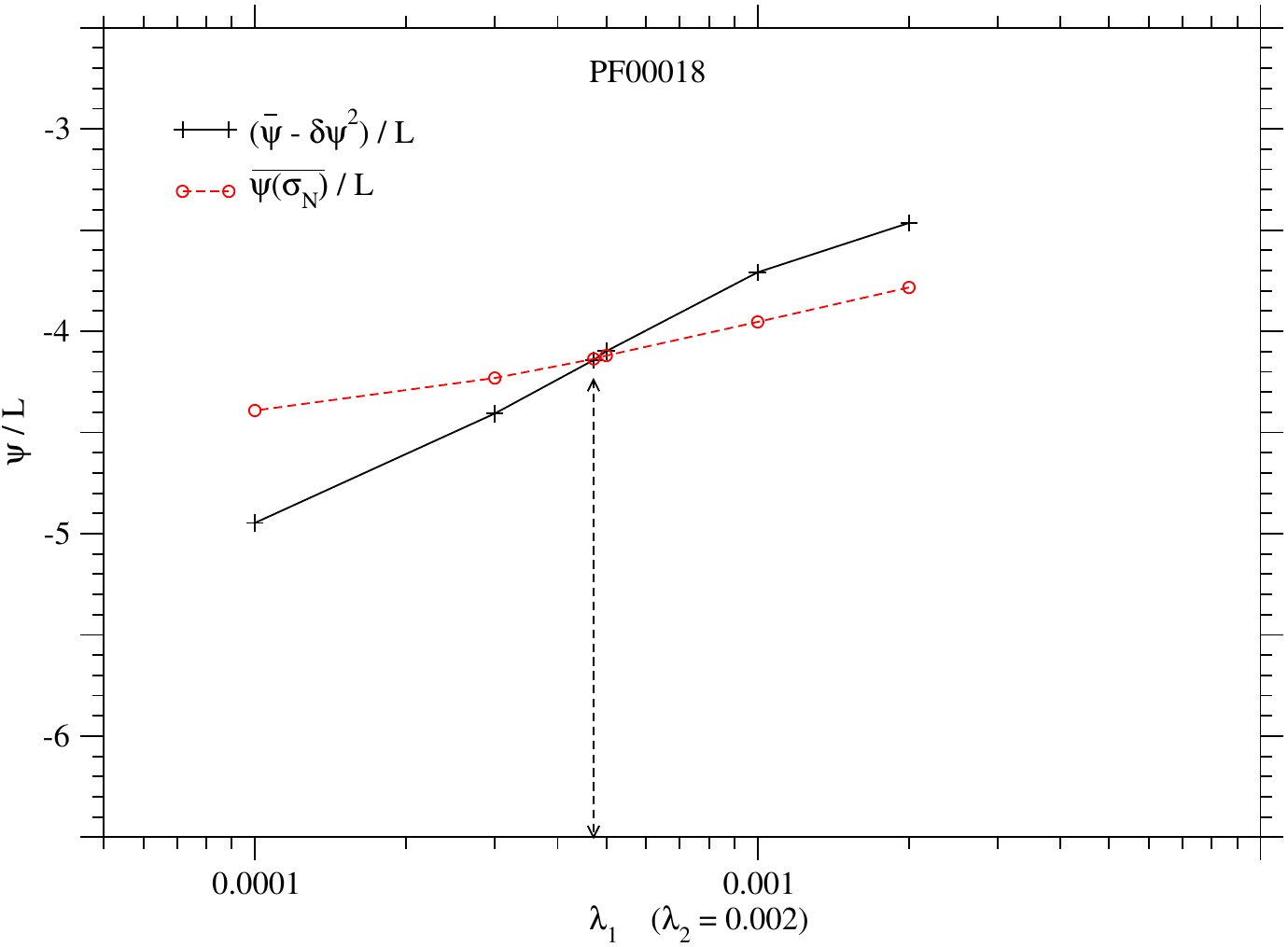}
}
\caption
{
\TEXTBF{How to adjust the regularization parameters $\lambda_1$ and $\lambda_2$ for \PRT;}
to take account of the gauge invariance of interactions, the Ising gauge that satisfies \Eq{\ref{\EQ: Ising gauge}} is employed.
\label{\FIG: tuning}
}
\end{figure*}
% End of Figs_prt/fig_tuning.tex
}% FigureInText

\TableInText{
\ifdefined\ThreeDigitsInTable
\else
\TCBB{
\newcommand{\ThreeDigitsInTable}[2]{#1}
}{
\newcommand{\ThreeDigitsInTable}[2]{#2}
}% TCBB
\fi

\begin{table}[!h]
\begin{threeparttable}[b]
\caption{
\TCBB{
\TEXTBF{
Hyperparameter values employed.
}
}{
\TEXTBF{
Hyperparameter values employed.
}
}% TCBB
}
\label{tbl: Parameter_range}
\newcommand{\ShowResults}[1]{}
\newcommand{\ShowParameters}[1]{#1}

\ifdefined\ShowParameters
\else
\newcommand{\ShowParameters}[1]{\Skip{#1}}
\fi

\ifdefined\ShowResults
\else
\newcommand{\ShowResults}[1]{\Skip{#1}}
\fi

\vspace*{1em}
\ShowResults{
\begin{tabular}{lrcccrccccc}
\hline
        & full-/mini- $^a$
        &       &
        & ModAdam & Adam
\\
}% ShowResults
\ShowParameters{
\begin{tabular}{lcccccccccccccc}
\hline
        & full-/mini- $^a$
        &       & &	& &      &
        & ModAdam &  Adam
\\
}% ShowParameters
Pfam ID
	& \, \, batch size
\ShowParameters{
	& $\lambda_{2:L}$ & & $\lambda_2$ & & $\lambda_{2:H}$ & $\lambda_1$
}% ShowParameters
\ShowResults{
	& $\lambda_2$ & $\lambda_1$		
}% ShowResults
	& $\kappa_{\text{max}}$ / $t_{\text{learning}}$  & $\kappa_{\text{max}}$ / $t_{\text{learning}}$ / decay-steps
\ShowResults{
	& $D^{KL}_2$
                & $\delta\psi^2 / L$ $^c$
                & $(\bar{\psi} - \delta\psi^2) / L$ $^{c,d}$
                & $\overline{\psi(\VEC{\sigma}_N)} / L$ $^e$
                & Precision $^g$
}% ShowResults
        \\
\hline
\hline
PF00018
	& 7007 / 103 
\ShowParameters{
	& 1.6e-3 &$\leq$ & 2.0e-3 & &  (6.0e-2$<$)   & 4.72e-4
}% ShowParameters{
\ShowResults{
	& 2.0e-3 & 4.72e-4
}% ShowResults
	& 0.006 / 1836 & 0.003 / 1836 / 2379
\ShowResults{
	& 0.0339487
	& 3.919889 & -4.098141 & -4.120034
	& 0.592105
}% ShowResults
	\\
PF00127
	& 2629 / 101
\ShowParameters{
	& 3.73e-3 & $\leq$ & 4.0e-3 & & (6.0e-2$<$)  & 4.0e-4
}% ShowParameters{
\ShowResults{
	& 4.0e-3 & 4.0e-4
}% ShowResults
	& 0.003 / 1872 & 0.002 / 1872 / 2287
\ShowResults{
	& 0.0482255
	& 4.261888 & -4.922126 & -4.921888
	& 0.459790
}% ShowResults
	\\
PF00153
	& 22061 / 105
\ShowParameters{
	& 1.15e-3 & $\leq$ & 1.4e-3 & & (1.0e-2$<$)  & 2.64e-4
}% ShowParameters{
\ShowResults{
	& 1.4e-3 & 2.64e-4
}% ShowResults
	& 0.006 / 1890 & 0.003 / 1890 / 2309
\ShowResults{
	& 0.0173416
	& 3.507434 & -3.902488 & -3.886850
	& 0.662651
}% ShowResults
	\\
PF00290
	& 3762 / 107
\ShowParameters{
	& 0.725e-3 &$\leq$ & 1.0e-3 &$<$ & 2.0e-3       & 1.52e-4
}% ShowParameters{
\ShowResults{
	& 1.0e-3 & 1.52e-4
}% ShowResults
	& 0.003 / 1890 & 0.0006 / 1890 / 6929 
\ShowResults{
	& 0.0185862 
	& 5.158079 & -5.651872 & -5.666304
	& 0.485157
}% ShowResults
	\\
PF00565
	& 9451 / 105
\ShowParameters{
	& 1.62e-3  &$\leq$ & 1.70e-3 &$<$ & 3.0e-3      & 3.5e-4
}% ShowParameters{
\ShowResults{
	& 1.7e-3 & 3.5e-4	
}% ShowResults
	& 0.010 / 1800 & 0.005 / 1800 / 2429
\ShowResults{
	& 0.0357605
	& 4.193700 & -4.532169 & -4.526433
	& 0.603406
}% ShowResults
	\\
PF00595
	& 5149 / 105
\ShowParameters{
	& 2.83e-3  &$\leq$ & 3.0e-3 & &   (6.0e-2$<$)   & 5.0e-4
}% ShowParameters{
\ShowResults{
	& 3.0e-3 & 5.0e-4
}% ShowResults
	& 0.006 / 1862 & 0.003 / 1862 / 2351
\ShowResults{
	& 0.0422348 
	& 3.952142 & -4.367744 & -4.363259
	& 0.573864
}% ShowResults
	\\
PF00887
	& 1965 / 109
\ShowParameters{
	& 1.84e-3 &$\leq$ & 2.0e-3 & $<$ & 6.0e-3     & 2.78e-4
}% ShowParameters{
\ShowResults{
	& 2.0e-3 & 2.78e-4
}% ShowResults
	& 0.006 / 1836 & 0.003 / 1836 / 2375 
\ShowResults{
	& 0.0365108
	& 5.637675 & -5.959571 & -5.920428
	& 0.553571
}% ShowResults
	\\
PF00959
	& 1564 / 104
\ShowParameters{
	& 0.538e-3 & $<$ & 1.07e-3 &$\leq$ & 1.2e-3   & 0.6e-7
}% ShowParameters{
\ShowResults{
	& 1.07e-3 & 0.6e-7 
}% ShowResults
	& 0.005 / 1890 & 0.002 / 1890 / 2519
\ShowResults{
	& 0.0219989
	& 5.723551 & -6.339924 & -6.289883
	& 0.414474
}% ShowResults
	\\
\hline
\end{tabular}
% End of Tables/parameter_range.tex
\begin{tablenotes}
\item [$^a$] The full-batch size is taken to be the number of representative sequences shown
in \Table{\ref{tbl: MSAs}}.
The mini-batch size is the minimum number of sequences employed to estimate the ensemble averages of pairwise amino acid frequencies
by the MCMC method at each learning step in the Boltzmann machine. 

% End of Tables/parameter_range_footnotes_3p.tex
\end{tablenotes}
\end{threeparttable}
\end{table}
% End of table_parameter_range.tex
}% TableInText

% End of Strategy_to_get_best.tex

% \input{results.tex}

\FigureInText{
\ifdefined\PRT
\renewcommand{\PRT}{PF00018}
\else
\newcommand{\PRT}{PF00018}
\fi
\ifdefined\DIR
\renewcommand{\DIR}[1]{Figs_prt/#1}
\else
\newcommand{\DIR}[1]{Figs_prt/#1}
\fi
\ifdefined\SUBDIR
\renewcommand{\SUBDIR}[1]{\DIR{\PRT/#1}}
\else
\newcommand{\SUBDIR}[1]{\DIR{\PRT/#1}}
\fi
\ifthenelse{\equal{\PRT}{PF00018}}{%

\begin{figure*}[!ht]
\centerline{
\includegraphics[width=114mm,angle=0]{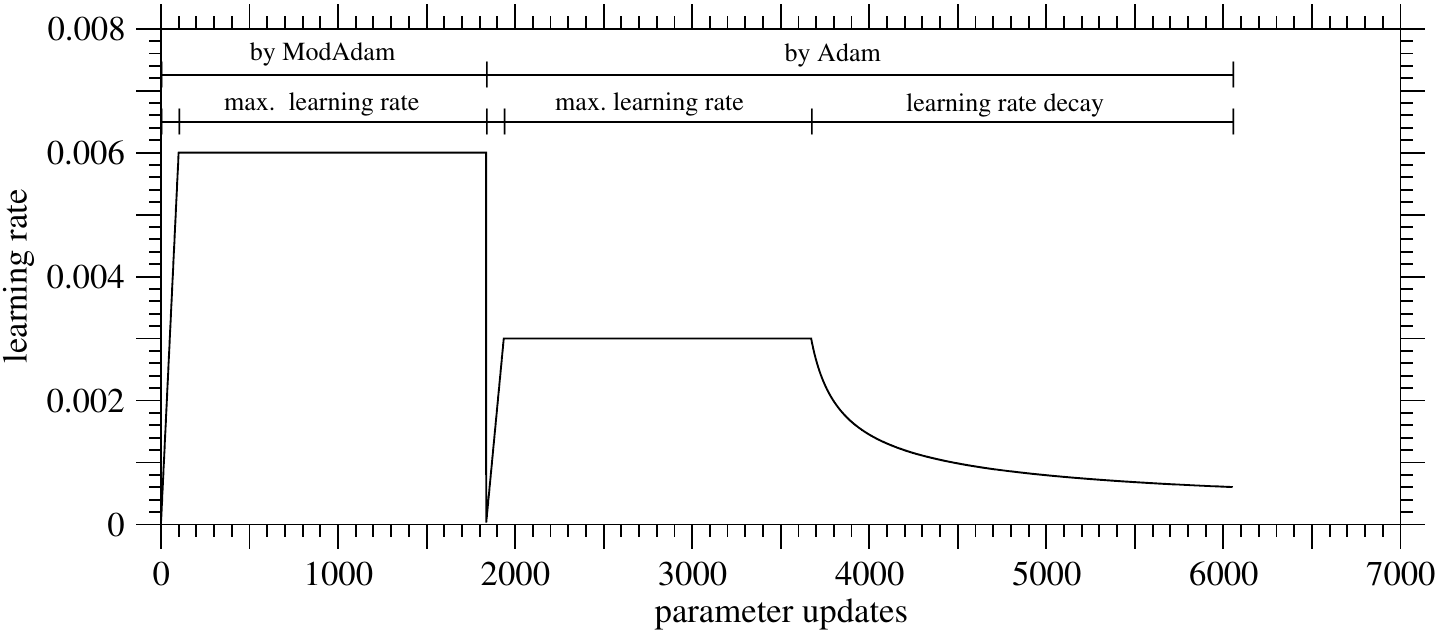}
}
\centerline{
\hspace*{3pt}
\includegraphics[width=112mm,angle=0]{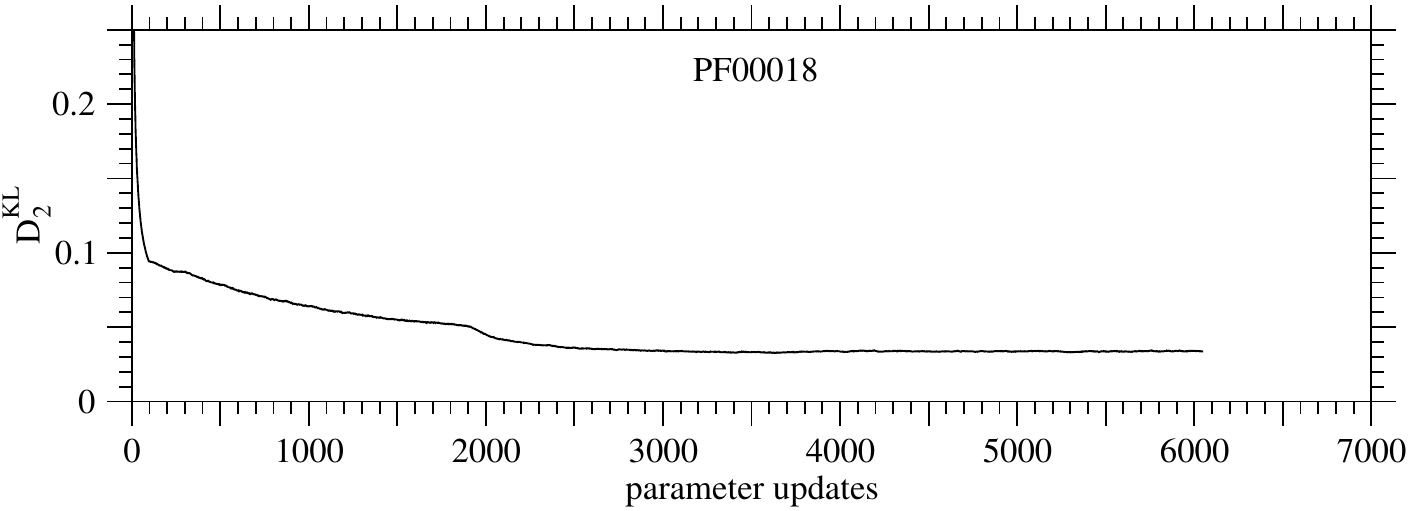}
}
\centerline{
\hspace*{7pt}
\includegraphics[width=110.5mm,angle=0]{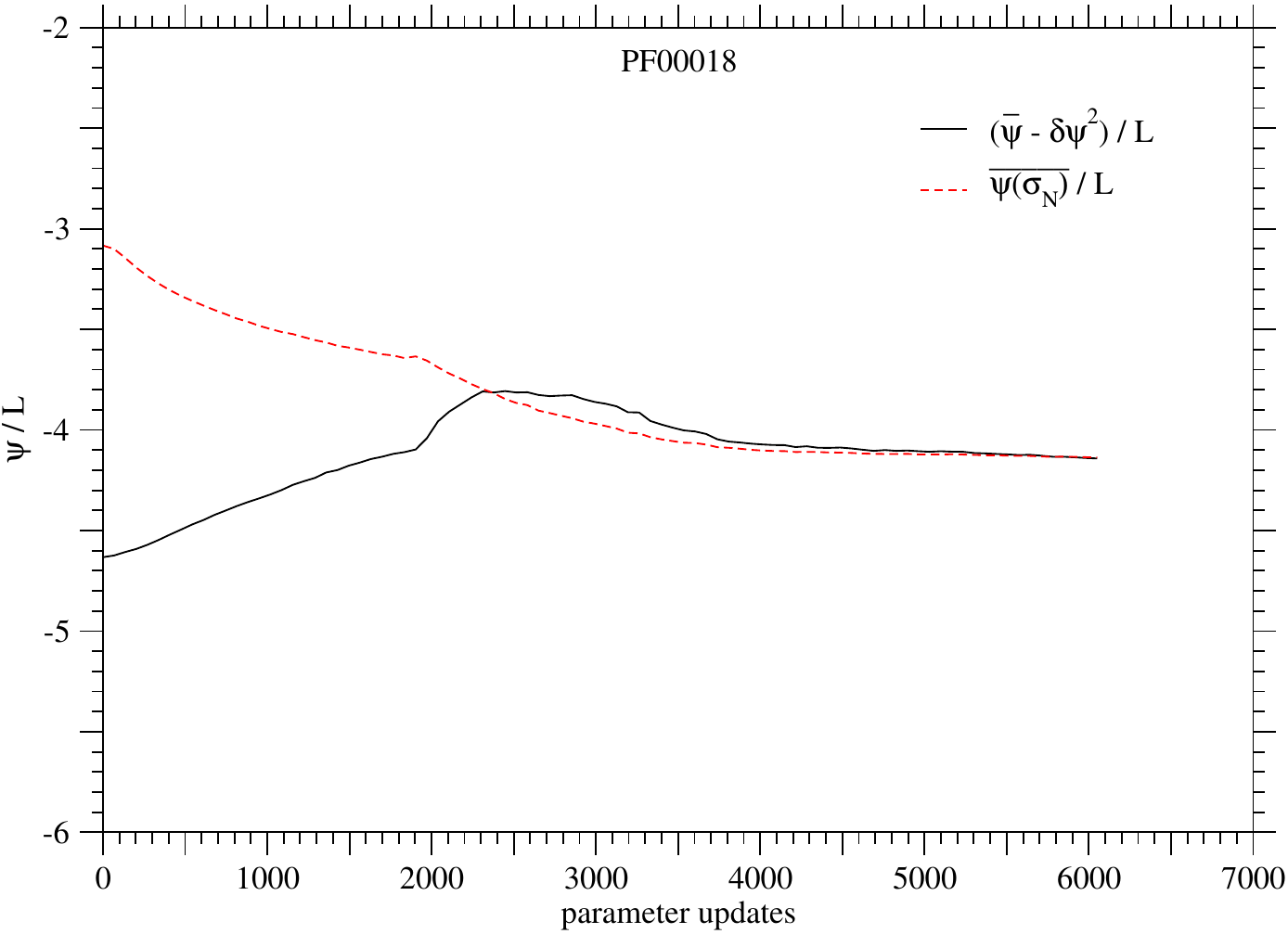}
}
% End of Figs_prt/PF00018/fig_prt.tex
}{% else
}
\caption
{
\TEXTBF{
For \PRT, the learning rate $\kappa(t)$, $D_2^{\text{KL}}$,
and $(\bar{\psi} - \delta\psi^2) / L$ and $\overline{\psi(\VEC{\sigma}_N)} / L$ at each step of 
the Boltzmann machine learning;}
to take account of the gauge invariance of interactions, the Ising gauge that satisfies \Eq{\ref{\EQ: Ising gauge}} is employed.
\label{\FIG: \PRT}
}
% End of Figs_prt/fig_prt_caption.tex
\end{figure*}
% End of Figs_prt/fig_prt.tex
}% FigureInText

\TableInText{
\ifdefined\ThreeDigitsInTable
\else
\TCBB{
\newcommand{\ThreeDigitsInTable}[2]{#1}
}{
\newcommand{\ThreeDigitsInTable}[2]{#2}
}% TCBB
\fi

\begin{table}[!h]
\begin{threeparttable}[b]
\caption{
\TCBB{
\TEXTBF{
Estimated values for each protein family
}
}{
\TEXTBF{
Estimated values for each protein family
}
}% TCBB
}
\label{tbl: Estimated_values}
\vspace*{1em}
\begin{tabular}{ccccccc}
\hline
\\
Pfam ID
	& $D^{KL}_1$ $^a$& $D^{KL}_2$ $^b$
                & $\delta\psi^2 / L$ $^c$
                & $(\bar{\psi} - \delta\psi^2) / L$ $^{c,d}$
                & $\overline{\psi(\VEC{\sigma}_N)} / L$ $^e$
                & Precision $^g$
        \\
\hline
\hline
PF00018
	& 0.002316   & 0.03364  		
	& 3.962    & -4.142    & -4.136   	
	& 0.612   				
	\\
PF00127
	& 0.001728   & 0.04823  		
	& 4.262    & -4.922    & -4.922   	
	& 0.460					
	\\
PF00153
	& 0.0008292   & 0.01736  		
	& 3.483    & -3.875    & -3.874   	
	& 0.663   				
	\\
PF00290
	& 0.001439   & 0.01809  		
	& 5.178    & -5.687    & -5.683   	
	& 0.499   				
	\\
PF00565
	& 0.001152   & 0.03576  		
	& 4.194    & -4.532    & -4.526   	
	& 0.603   				
	\\
PF00595
	& 0.002006   & 0.04223  		
	& 3.952    & -4.368    & -4.363   	
	& 0.574   				
	\\
PF00887
	& 0.001861   & 0.03630  		
	& 5.571    & -5.887    & -5.895   	
	& 0.539   				
	\\
PF00959
	& 0.0008958   & 0.02479  		
	& 5.484    & -6.137    & -6.132   	
	& 0.445   				
	\\
\hline
\end{tabular}
% End of Tables/outcome.tex
\begin{tablenotes}

\item [$^a$] 
$D_1^{KL}$ is the average over all residue positions of Kullback-Leibler divergences for single-site amino acid frequencies
between the reference and model;
about 10000 MCMC samples of multiple mini-batches 
in the last learning steps are employed to estimate
amino acid frequencies at each residue position. 

\item [$^b$] 
$D_2^{KL}$ is the average over all residue pairs of Kullback-Leibler divergences for pairwise marginal distributions
between the reference and model;
about 10000 MCMC samples of multiple mini-batches 
in the last learning steps are employed to estimate
pairwise amino acid frequencies at each site pair. 

\item [$^c$] $\bar{\psi}$ and $\delta \psi^2$ are the mean and variance of the total interactions $\psi(\VECS{\sigma})$ of random sequences
whose amino acid composition is one calculated by \Eq{\ref{\EQ: Pi_for_BM}}; the Ising gauge is employed.

\item [$^d$] The ensemble average of the total interactions per residue, $\langle \psi(\VECS{\sigma}) \rangle_{\VECS{\sigma}} / L$, in the Boltzmann distribution 
by the Gaussian approximation for the $\psi$ density.
The mean and variance of the Gaussian distribution are evaluated as ones of $\psi(\VECS{\sigma})$ of random sequences. 

\item [$^e$] The average of the total interactions per residue over the representatives of the native sequences, 
which are more than 20\% different from any other; the Ising gauge is employed.

\item [$^g$] Precision of contact residue pair prediction; the total number of predicted contacts is
equal to the total number of closely located residue pairs
within $8$ \AA\  between side-chain centers in the 3D protein structure.
The corrected Frobenius norm of couplings
is employed for the contact score\CITE{ELLWA:13,EHA:14}.

% End of Tables/outcome_footnotes_3p.tex
\end{tablenotes}
\end{threeparttable}
\end{table}
% End of table_outcome.tex
}% TableInText

\SECTION{Results}

In \Table{\ref{\TBL: Estimated_values}},
the estimated values of 
$\bar{\psi} - \delta \psi^2$ and $\overline{\psi(\VEC{\sigma}_N)}$
are shown in the Ising gauge
as well as others for each protein family.
\Fig{\ref{\FIG: \PRT}} shows 
the learning profile in the Ising gauge for \PRT,
which consists of the learning schedule, and the changes of the average Kullback-Leibler divergence
of pairwise marginal distributions over all residue pairs,
the mean of the total interactions $\overline{\psi(\VEC{\sigma}_N)}$
over the representatives of the native proteins,
and the ensemble average $\langle \psi(\VEC{\sigma}) \rangle_{\VEC{\sigma}} = \bar{\psi} - \delta \psi^2$.
The learning profiles for the other proteins
are shown in the supplement.

As learning steps progress, the $D_2^{KL}$ exhibits a smooth
downward trend without indicating anything wrong.
Here it should be noticed that 
the fluctuations of
the $D_1^{KL}$ and $D_2^{KL}$
are reduced 
by employing multiple mini-batches in the preceding steps,
which correspond to about 10000 samples.
If $D_1^{KL}$ and $D_2^{KL}$ are estimated for single mini-batches,
they will somewhat fluctuate in the learning process. 
The pairwise marginal distributions
estimated from the small size of mini-batch by MCMC samplings
would include significant statistical errors,
and therefore the partial derivatives of the objective function
evaluated from those estimated values also include statistical errors 
causing fluctuations of the learning

Both of 
the average of $\psi(\VEC{\sigma}_N)$ over all representatives of the native sequences
and the ensemble average $\langle \psi(\VEC{\sigma}) \rangle_{\VEC{\sigma}}$ change
in most cases
smoothly along with the learning steps.
The $\overline{\psi(\VEC{\sigma}_N)}$ exhibits a downward trend, 
On the other hand, the $\langle \psi(\VEC{\sigma}) \rangle_{\VEC{\sigma}}$
does not show such a simple trend,
although
both the $\overline{\psi(\VEC{\sigma})}$ and $\langle \psi(\VEC{\sigma}) \rangle_{\VEC{\sigma}}$
converge to the same value.
When the two regularization parameters $\lambda_1$ and $\lambda_2$, the maximum learning rate $\kappa_{\text{max}}$,
and decay steps are best adjusted,
these are common features shown for all the protein families studied here; see figures in the supplement.

% End of results.tex

% \input{disc.tex}

\SECTION{Discussion}

The reproducibilities of single-site frequencies and pairwise correlations by the Boltzmann machine method 
were confirmed\CITE{BLCC:16,CFFMW:17,FBW:18}  to be better than 
those by the approximate methods such as the mean field approximation\CITE{LGJ:02,LGJ:12,MPLBMSZOHW:11,MCSHPZS:11} and
pseudo-likelihood maximization method\CITE{ELLWA:13,EHA:14}.
The reproducibilities of fields and couplings
were examined mostly for artificial systems such as 
artificial data on Erd\"{o}s-Renyi models and 
a lattice protein\CITE{BLCC:16}.
Their reproducibilities for protein-like systems,
which were generated with the fields and couplings
estimated for protein families by the Boltzmann machine method, were studied\CITE{M:19,M:20};
both the fields and couplings were well reproduced
except for very weak couplings.

In order to find the appropriate estimates of fields and couplings for
protein families, the regularization parameters ($\lambda_1$ and $\lambda_2$), 
and the learning schedule (
the maximum learning rate $\kappa_{\max}$, the learning steps $t_{\text{learning}}$,
and decay steps ) must be optimized for protein families.
Here, the hyperparameters are adjusted to make 
the average, $\overline{\psi(\VEC{\sigma}_N)}$, of the total interactions over the native sequences 
to be equal to its ensemble average, $\langle \psi(\VEC{\sigma}) \rangle_\VEC{\sigma} = \bar{\psi} - \delta \psi^2$.
This condition for protein structures is indicated by the protein folding theories\CITE{SG:93a,SG:93b,RS:94,PGT:97}.
In addition, we make the hyperparameters to minimize $\overline{\psi(\VEC{\sigma}_N)} \simeq \bar{\psi} - \delta \psi^2$,
where the interactions are represented in the Ising gauge.

However, a computational load is very high for
the Boltzmann machine method to estimate fields and couplings.
Here we tried to reduce computational time by using the parallel, persistent MCMC 
to estimate 
the ensemble averages of single-site and pairwise amino acid frequencies
and by learning with the stochastic gradient descent methods.

Alternatively,
restricted Boltzmann machines that are equivalent to the present Boltzmann machine
have been studied\CITE{TCM:19,SW:19}. In these models,
the coupling interactions $J_{ij}(a_k, a_l)$ are estimated in the decoupled form, 
$\sum_{\mu=1}^{L_q} \xi_i^{\mu}(a_k) \xi_j^{\mu}(a_l)$, 
and approximated with the small numbers for $L_{q}$, reducing a computational load;
the number of parameters for coupling interactions 
is reduced from $L(L-1)q^2/2$ to $LL_{q}q$;
$q = 21$.
Thus, using the restricted Boltzmann machines 
certainly has a merit, although $L_{q}$
may not be much less than $O(Lq)$ to well approximate sparse coupling interactions,
which is the case for proteins.

% End of disc.tex

% \input{declarations.tex}

\vspace*{2em}
\noindent
\textit{The program written in Scala and the MSAs employed are available from
https://gitlab.com/sanzo.miyazawa/BM/
\newline
and
https://github.com/Sanzo-Miyazawa/BM/ .}

% End of declarations.tex

\clearpage

\TCBB{

}{

}% TCBB

}% TEXT

% End of contents_text.tex

%%%%%%%%%%%%%

\ifCLASSOPTIONcaptionsoff
  \newpage
\fi



\bibliographystyle{IEEEtran}
\bibliography{jnames_with_dots,SM,Protein,MolEvol,Bioinfo}
%

%\Skip{
%\begin{IEEEbiographynophoto}{Sanzo Miyazawa}
%\begin{IEEEbiography}[{\includegraphics[width=1in,height=1.25in,clip,keepaspectratio]{Sanzo_photo}}]{Sanzo Miyazawa}
%\begin{IEEEbiography}[{\includegraphics[height=0.65in,clip,keepaspectratio,angle=90]{Sanzo_photo}}]{Sanzo Miyazawa}
%\begin{IEEEbiography}[{\includegraphics[height=0.8in,clip,keepaspectratio,angle=90]{Sanzo_photo}}]{Sanzo Miyazawa}
%\Skip{
\begin{IEEEbiographynophoto}{Sanzo Miyazawa}
had worked for the Graduate School of Engineering in Gunma University, Japan until retired at age 65 in 2013.
His research interests include protein structure and evolution.
\end{IEEEbiographynophoto}
%}% Skip
%}% Skip


%\vfill


\TCBB{
% FiguresWithoutCaption
}{
}% TCBB

\renewcommand{\TEXT}[1]{}
\renewcommand{\SUPPLEMENT}[1]{#1}

\FiguresWithoutCaption{
\renewcommand{\SUPPLEMENT}[1]{}
}% FiguresWithoutCaption

\SUPPLEMENT{
\renewcommand{\SkipSupplToMerge}[1]{#1}
\SkipSupplToMerge{
%%%%%%%%%%
%\input{contents_supplement_TCBB.tex}
% \input{contents_supplement.tex}

\SUPPLEMENT{
\SkipSupplToMerge{

\renewcommand{\TEXT}[1]{}
\renewcommand{\SUPPLEMENT}[1]{#1}
\renewcommand{\TBL}{stbl}
\renewcommand{\FIG}{sfig}

\clearpage
\newpage

\arXiv{

\setcounter{page}{1}
\renewcommand{\thepage}{S-\arabic{page}}

}{
\setcounter{page}{1}
\renewcommand{\thepage}{S-\arabic{page}}
}% arXiv

\setcounter{section}{0}
\renewcommand{\thesection}{S.\arabic{section}}

\setcounter{equation}{0}
\renewcommand{\theequation}{S\arabic{equation}}

\setcounter{table}{0}
\renewcommand{\thetable}{S\arabic{table}}

\setcounter{figure}{0}
\renewcommand{\thefigure}{S\arabic{figure}}

\TCBB{

\begin{figure*}

\Large

\begin{center}
\textbf{Supplementary Material} \\
for \\
Boltzmann Machine learning with \\
a Parallel, Persistent Markov chain Monte Carlo method \\
for Estimating Evolutionary Fields and Couplings \\
from a Protein Multiple Sequence Alignment
\end{center}

\vspace*{1em}
\begin{center}
Sanzo Miyazawa  \\
sanzo.miyazawa@gmail.com
\end{center}
\begin{center}
2026-04-19
\end{center}

\normalsize

\end{figure*}
% End of cover.tex
\clearpage
\newpage

\SECTION{Figures}

\TEXT{

\TCBB{

}{

}% TCBB

}{

\TCBB{
}{

}% TCBB

}% TEXT

\SkipSupplToMerge{
\SUPPLEMENT{

\ifdefined\DIR
\renewcommand{\DIR}[1]{Figs_prt/#1}
\else
\newcommand{\DIR}[1]{Figs_prt/#1}
\fi
\ifdefined\PRT
\renewcommand{\PRT}{PF00018}
\else
\newcommand{\PRT}{PF00018}
\fi
\ifdefined\DIR
\renewcommand{\DIR}[1]{Figs_prt/#1}
\else
\newcommand{\DIR}[1]{Figs_prt/#1}
\fi
\ifdefined\SUBDIR
\renewcommand{\SUBDIR}[1]{\DIR{\PRT/#1}}
\else
\newcommand{\SUBDIR}[1]{\DIR{\PRT/#1}}
\fi
\ifthenelse{\equal{\PRT}{PF00018}}{%

\begin{figure*}[!ht]
\centerline{
\includegraphics[width=114mm,angle=0]{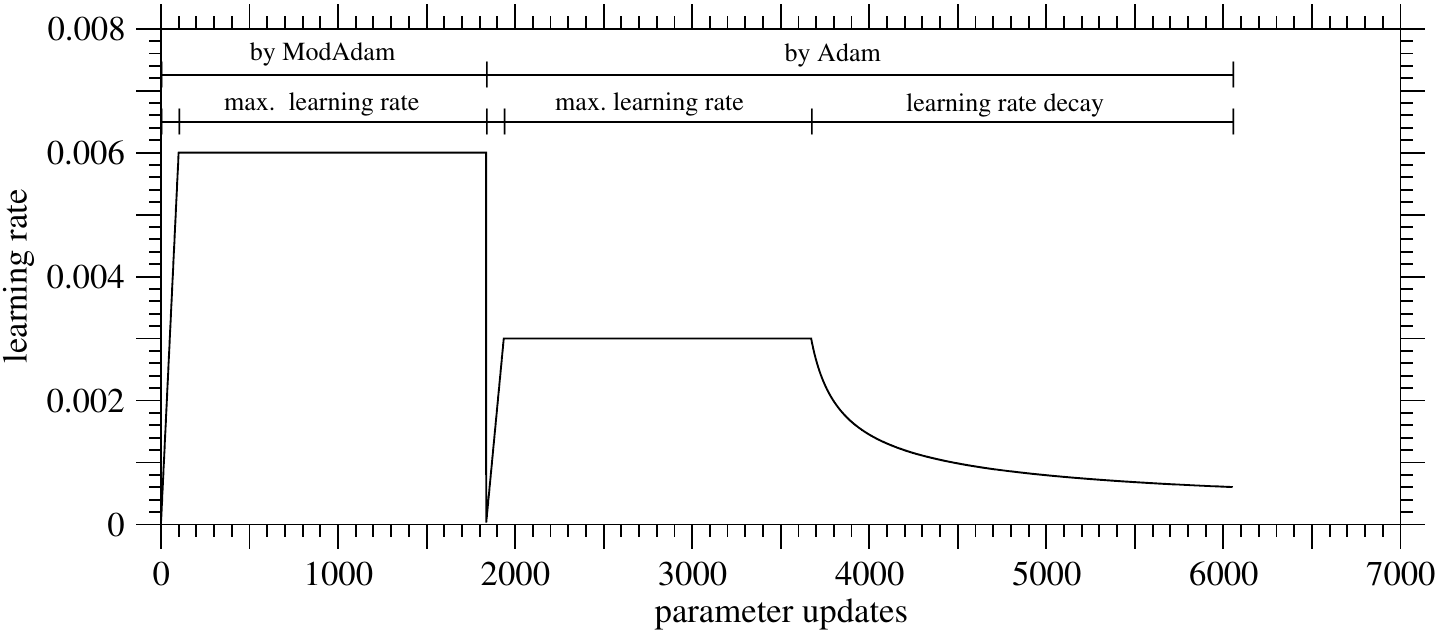}
}
\centerline{
\hspace*{3pt}
\includegraphics[width=112mm,angle=0]{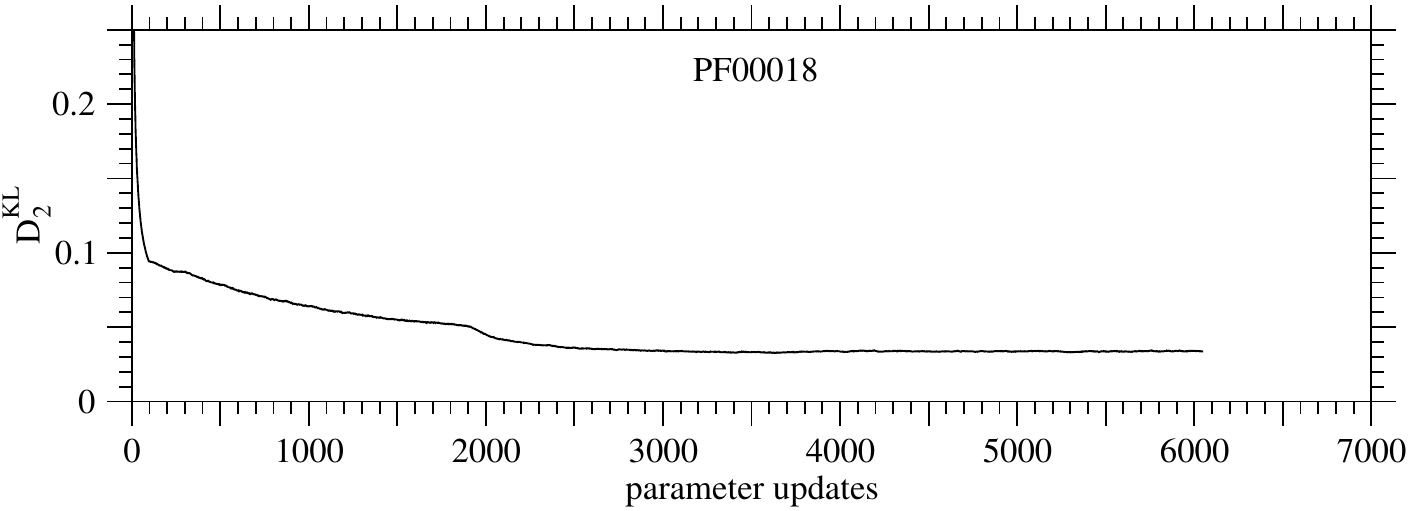}
}
\centerline{
\hspace*{7pt}
\includegraphics[width=110.5mm,angle=0]{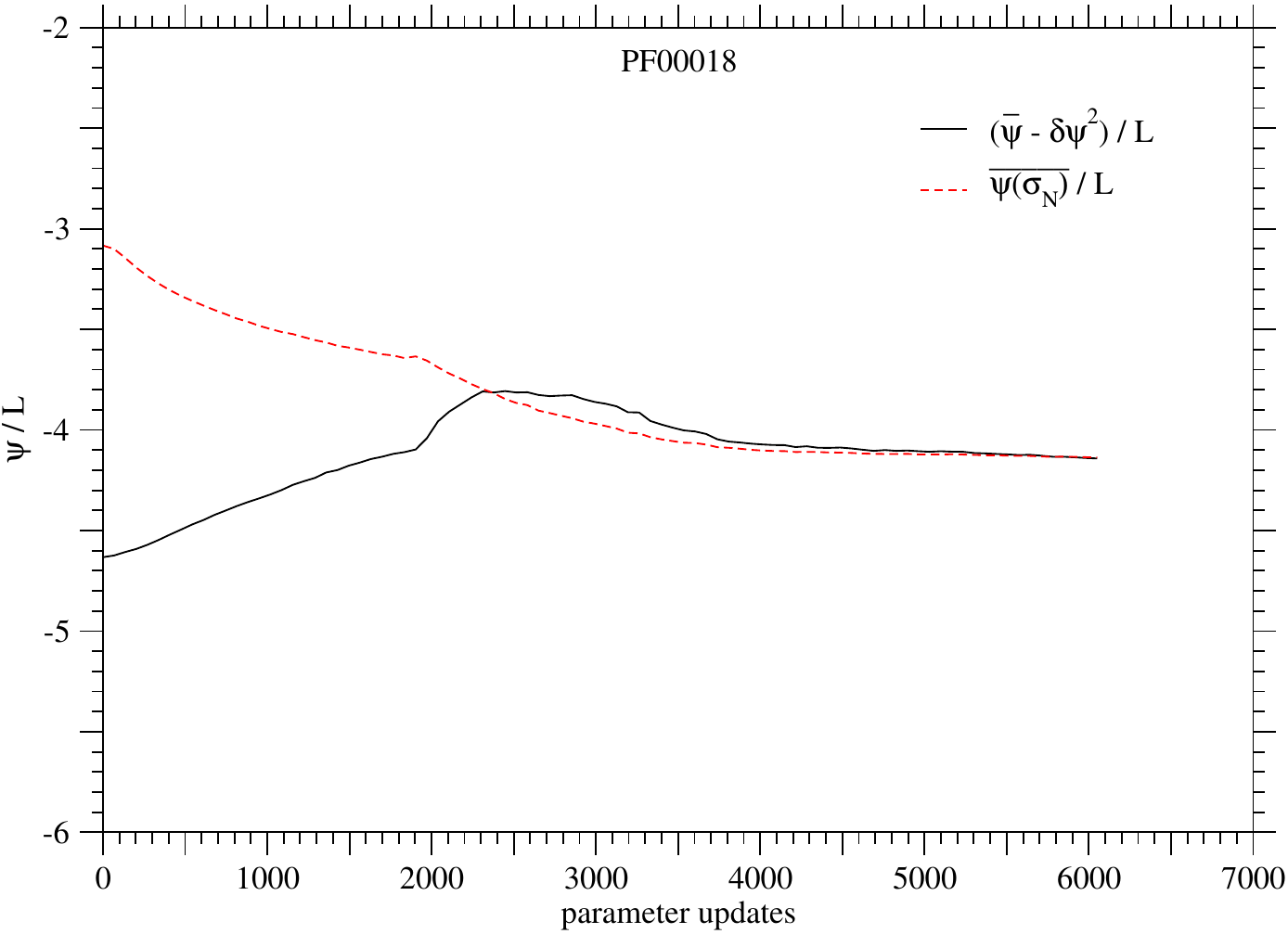}
}
% End of Figs_prt/PF00018/fig_prt.tex
}{% else
}
\caption
{
\TEXTBF{
For \PRT, the learning rate $\kappa(t)$, $D_2^{\text{KL}}$,
and $(\bar{\psi} - \delta\psi^2) / L$ and $\overline{\psi(\VEC{\sigma}_N)} / L$ at each step of 
the Boltzmann machine learning;}
to take account of the gauge invariance of interactions, the Ising gauge that satisfies \Eq{\ref{\EQ: Ising gauge}} is employed.
\label{\FIG: \PRT}
}
% End of Figs_prt/fig_prt_caption.tex
\end{figure*}
% End of Figs_prt/fig_prt_pf00018.tex

\renewcommand{\PRT}{PF00127}
\ifdefined\DIR
\renewcommand{\DIR}[1]{Figs_prt/#1}
\else
\newcommand{\DIR}[1]{Figs_prt/#1}
\fi
\ifdefined\SUBDIR
\renewcommand{\SUBDIR}[1]{\DIR{\PRT/#1}}
\else
\newcommand{\SUBDIR}[1]{\DIR{\PRT/#1}}
\fi
\ifthenelse{\equal{\PRT}{PF00127}}{%

\begin{figure*}[!ht]
\centerline{
\includegraphics[width=114mm,angle=0]{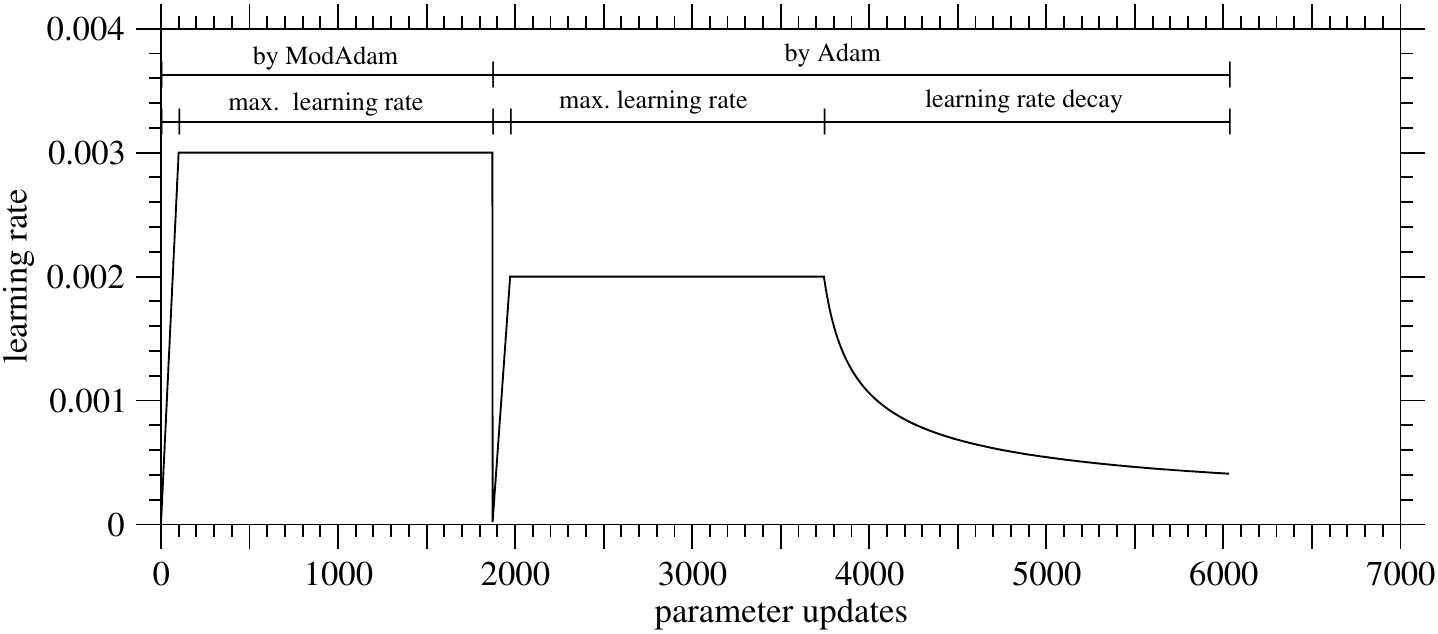}
}
\centerline{
\hspace*{3pt}
\includegraphics[width=112mm,angle=0]{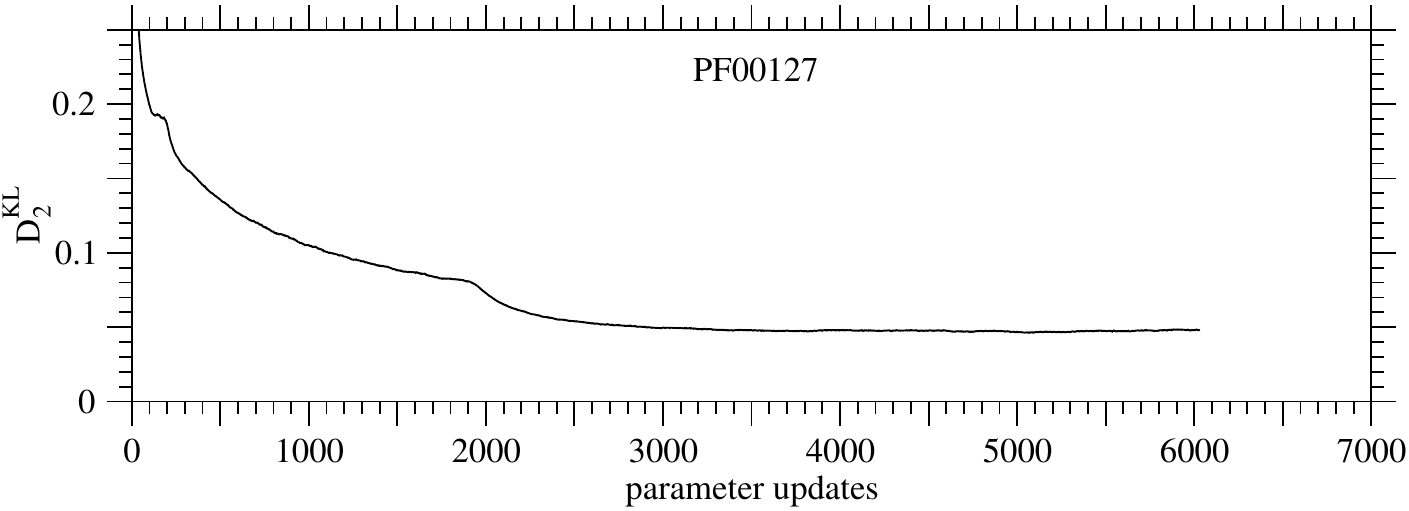}
}
\centerline{
\hspace*{7pt}
\includegraphics[width=110.5mm,angle=0]{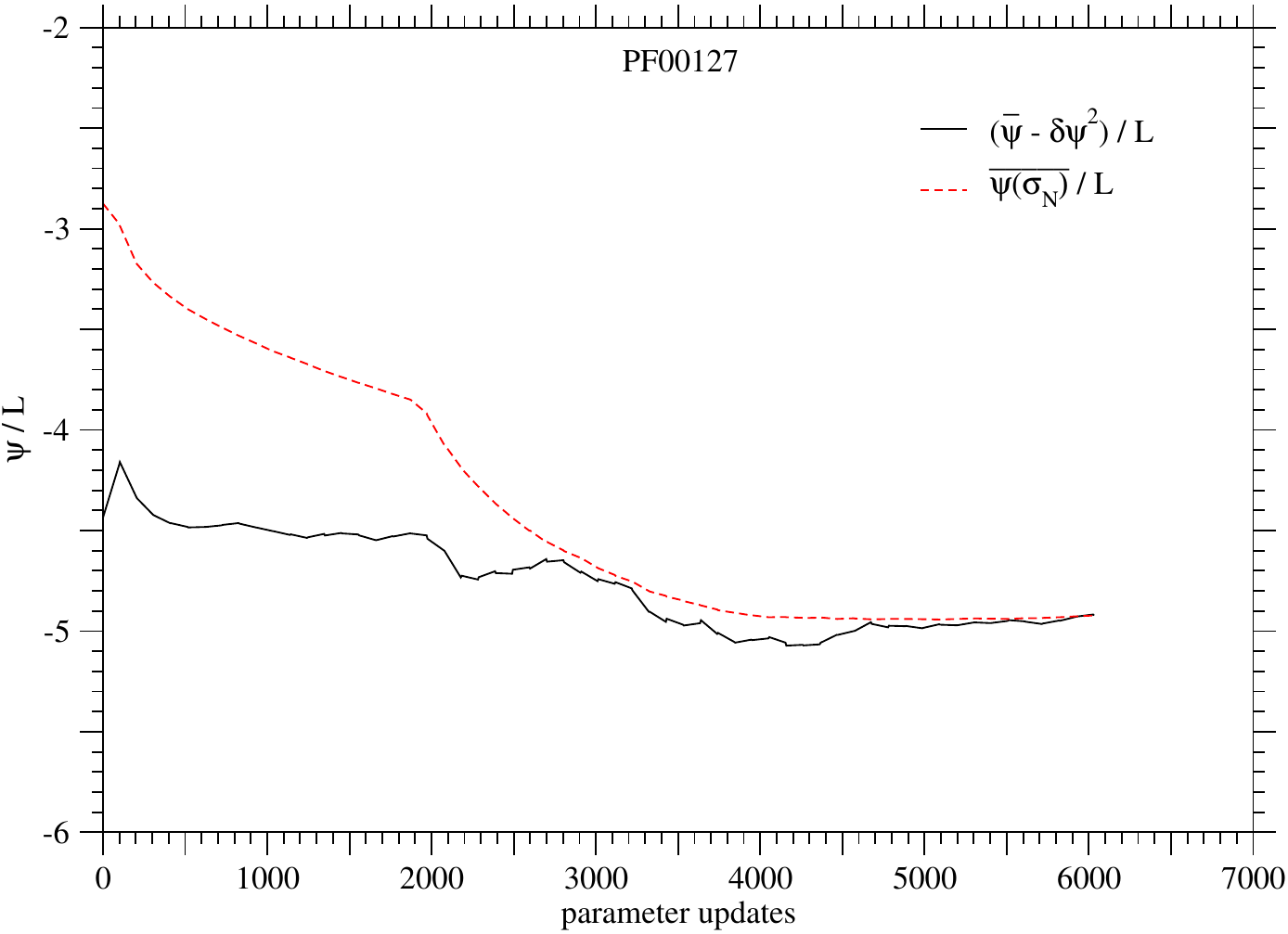}
}
% End of Figs_prt/PF00127/fig_prt.tex
}{% else
}
\caption
{
\TEXTBF{
For \PRT, the learning rate $\kappa(t)$, $D_2^{\text{KL}}$,
and $(\bar{\psi} - \delta\psi^2) / L$ and $\overline{\psi(\VEC{\sigma}_N)} / L$ at each step of 
the Boltzmann machine learning;}
to take account of the gauge invariance of interactions, the Ising gauge that satisfies \Eq{\ref{\EQ: Ising gauge}} is employed.
\label{\FIG: \PRT}
}
% End of Figs_prt/fig_prt_caption.tex
\end{figure*}
% End of Figs_prt/fig_prt_pf00127.tex

\renewcommand{\PRT}{PF00153}
\ifdefined\DIR
\renewcommand{\DIR}[1]{Figs_prt/#1}
\else
\newcommand{\DIR}[1]{Figs_prt/#1}
\fi
\ifdefined\SUBDIR
\renewcommand{\SUBDIR}[1]{\DIR{\PRT/#1}}
\else
\newcommand{\SUBDIR}[1]{\DIR{\PRT/#1}}
\fi
\ifthenelse{\equal{\PRT}{PF00153}}{%

\begin{figure*}[!ht]
\centerline{
\includegraphics[width=114mm,angle=0]{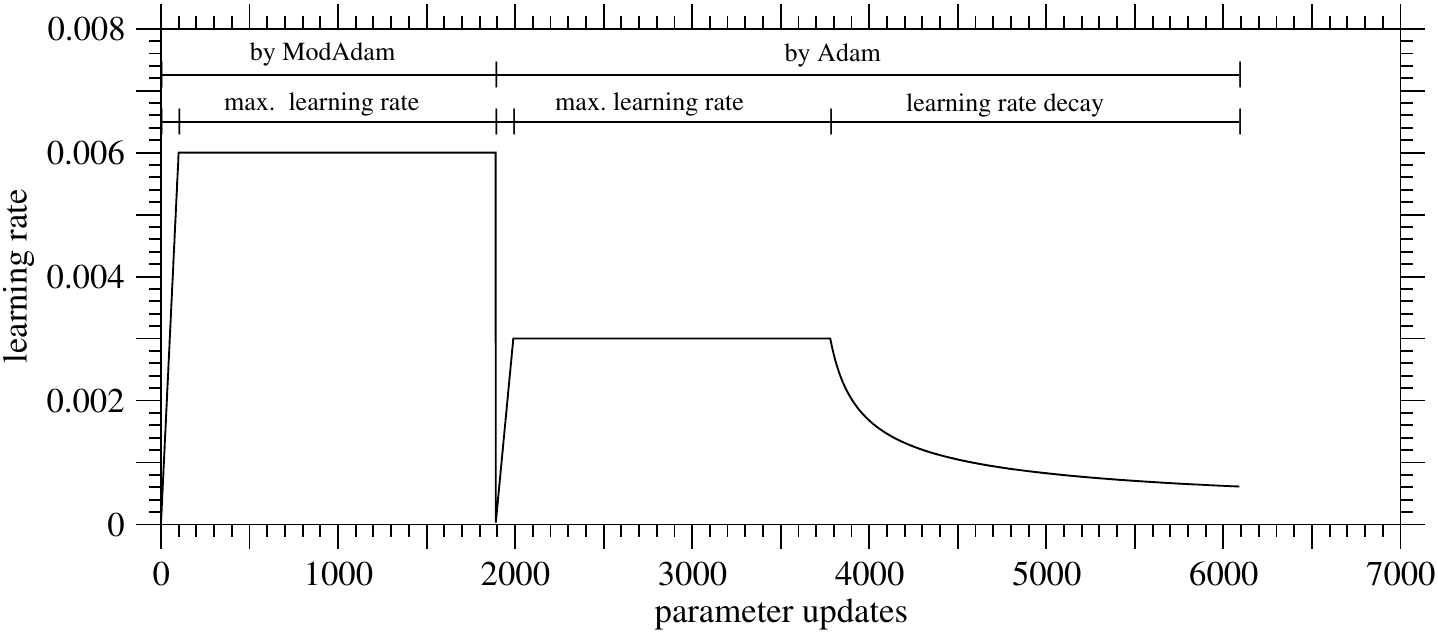}
}
\centerline{
\hspace*{3pt}
\includegraphics[width=112mm,angle=0]{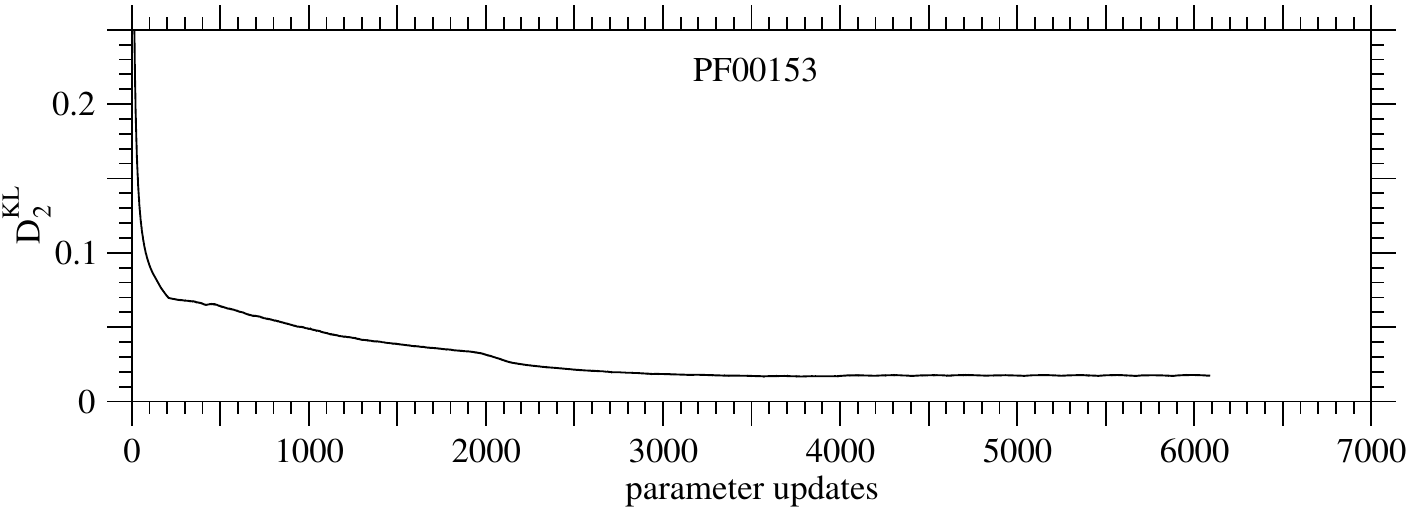}
}
\centerline{
\hspace*{7pt}
\includegraphics[width=110.5mm,angle=0]{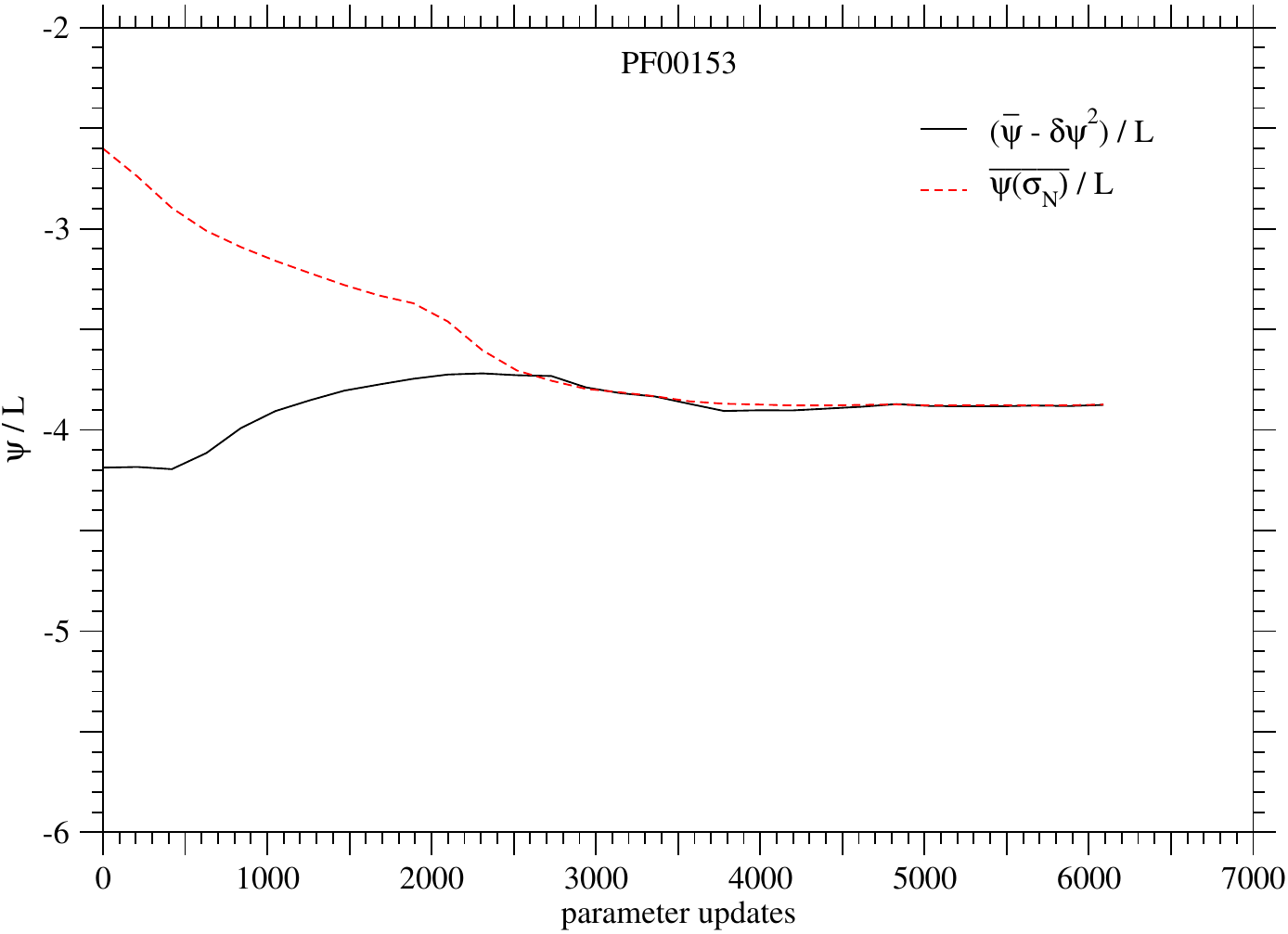}
}
% End of Figs_prt/PF00153/fig_prt.tex
}{% else
}
\caption
{
\TEXTBF{
For \PRT, the learning rate $\kappa(t)$, $D_2^{\text{KL}}$,
and $(\bar{\psi} - \delta\psi^2) / L$ and $\overline{\psi(\VEC{\sigma}_N)} / L$ at each step of 
the Boltzmann machine learning;}
to take account of the gauge invariance of interactions, the Ising gauge that satisfies \Eq{\ref{\EQ: Ising gauge}} is employed.
\label{\FIG: \PRT}
}
% End of Figs_prt/fig_prt_caption.tex
\end{figure*}
% End of Figs_prt/fig_prt_pf00153.tex

\renewcommand{\PRT}{PF00290}
\ifdefined\DIR
\renewcommand{\DIR}[1]{Figs_prt/#1}
\else
\newcommand{\DIR}[1]{Figs_prt/#1}
\fi
\ifdefined\SUBDIR
\renewcommand{\SUBDIR}[1]{\DIR{\PRT/#1}}
\else
\newcommand{\SUBDIR}[1]{\DIR{\PRT/#1}}
\fi
\ifthenelse{\equal{\PRT}{PF00290}}{%

\begin{figure*}[!ht]
\centerline{
\includegraphics[width=114mm,angle=0]{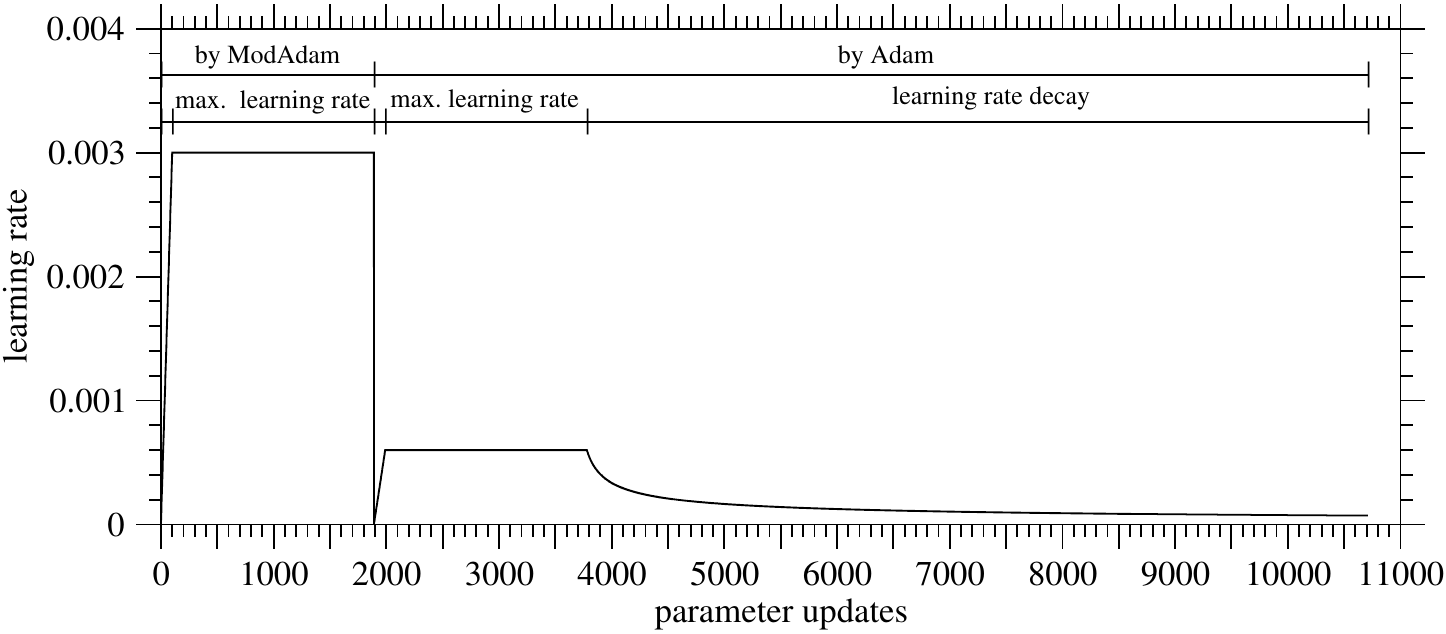}
}
\centerline{
\hspace*{3pt}
\includegraphics[width=112mm,angle=0]{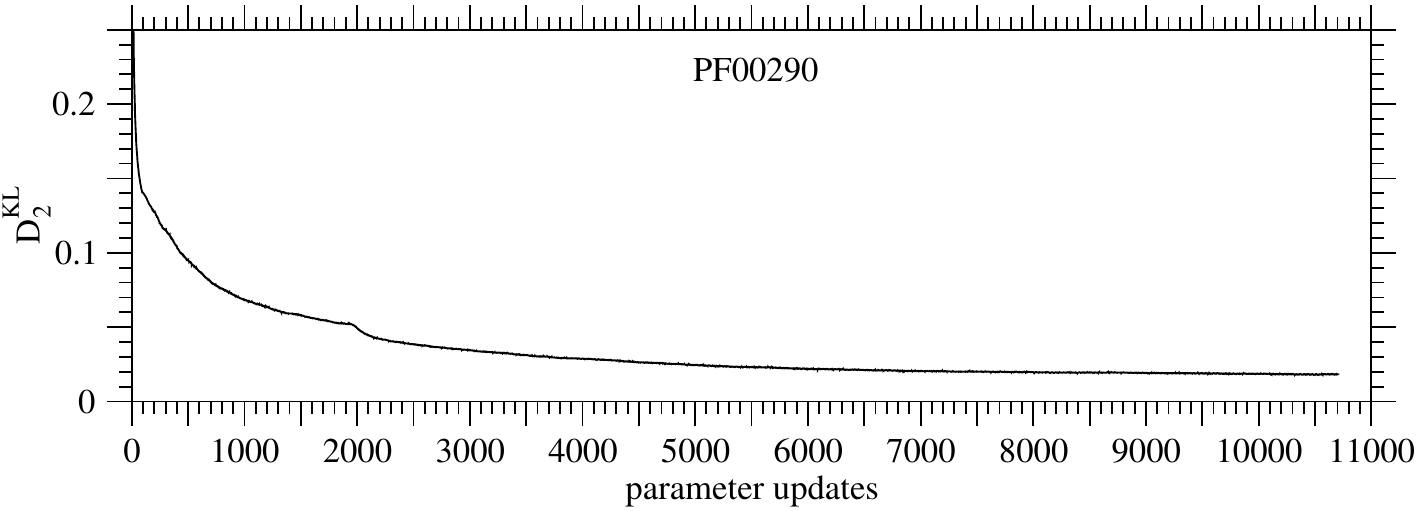}
}
\centerline{
\hspace*{7pt}
\includegraphics[width=110.5mm,angle=0]{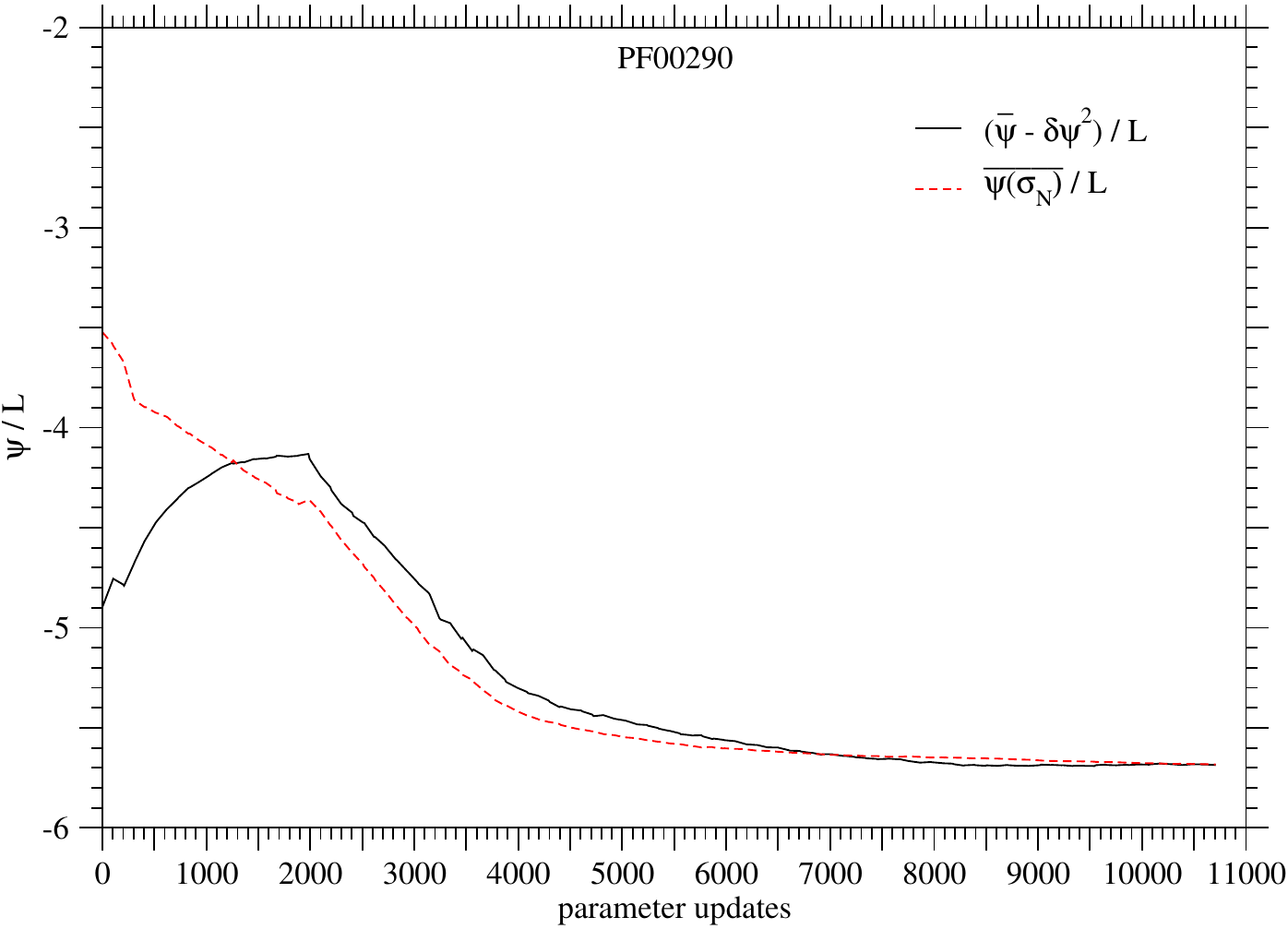}
}
% End of Figs_prt/PF00290/fig_prt.tex
}{% else
}
\caption
{
\TEXTBF{
For \PRT, the learning rate $\kappa(t)$, $D_2^{\text{KL}}$,
and $(\bar{\psi} - \delta\psi^2) / L$ and $\overline{\psi(\VEC{\sigma}_N)} / L$ at each step of 
the Boltzmann machine learning;}
to take account of the gauge invariance of interactions, the Ising gauge that satisfies \Eq{\ref{\EQ: Ising gauge}} is employed.
\label{\FIG: \PRT}
}
% End of Figs_prt/fig_prt_caption.tex
\end{figure*}
% End of Figs_prt/fig_prt_pf00290.tex

\renewcommand{\PRT}{PF00565}
\ifdefined\DIR
\renewcommand{\DIR}[1]{Figs_prt/#1}
\else
\newcommand{\DIR}[1]{Figs_prt/#1}
\fi
\ifdefined\SUBDIR
\renewcommand{\SUBDIR}[1]{\DIR{\PRT/#1}}
\else
\newcommand{\SUBDIR}[1]{\DIR{\PRT/#1}}
\fi
\ifthenelse{\equal{\PRT}{PF00565}}{%

\begin{figure*}[!ht]
\centerline{
\includegraphics[width=114mm,angle=0]{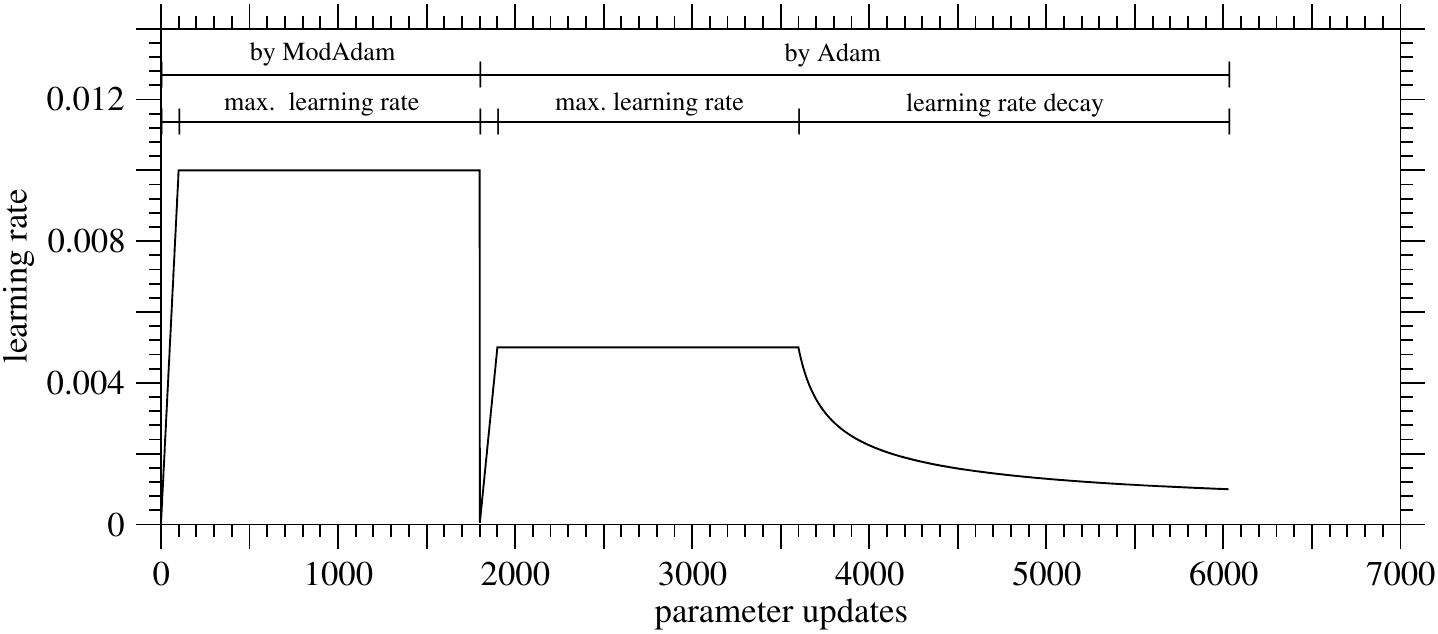}
}
\centerline{
\hspace*{3pt}
\includegraphics[width=112mm,angle=0]{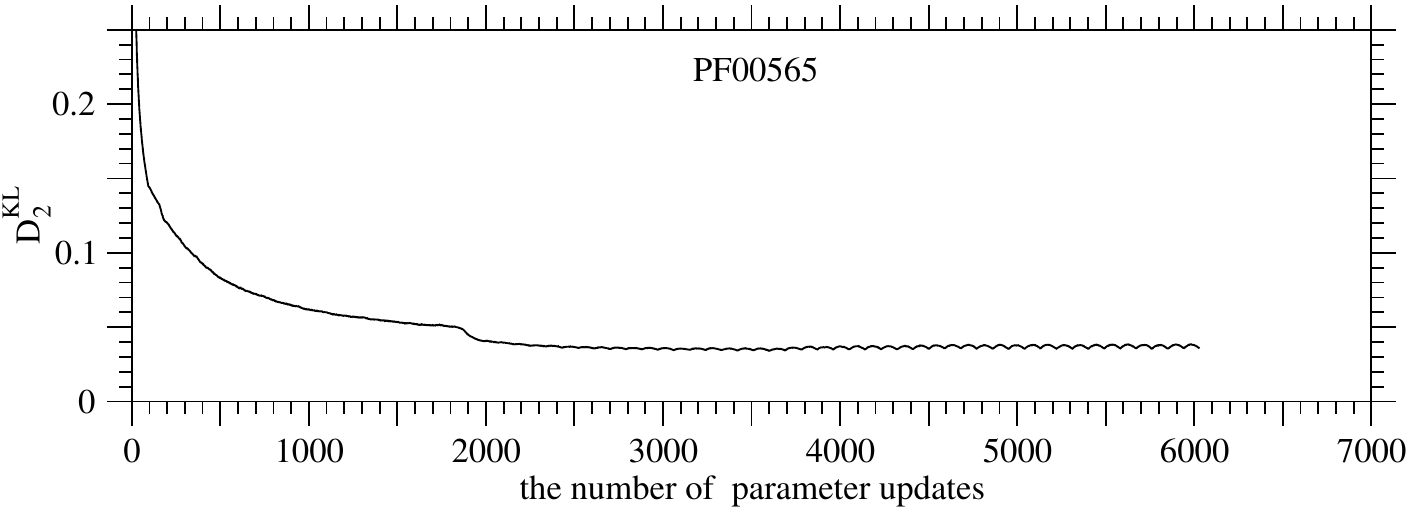}
}
\centerline{
\hspace*{7pt}
\includegraphics[width=110.5mm,angle=0]{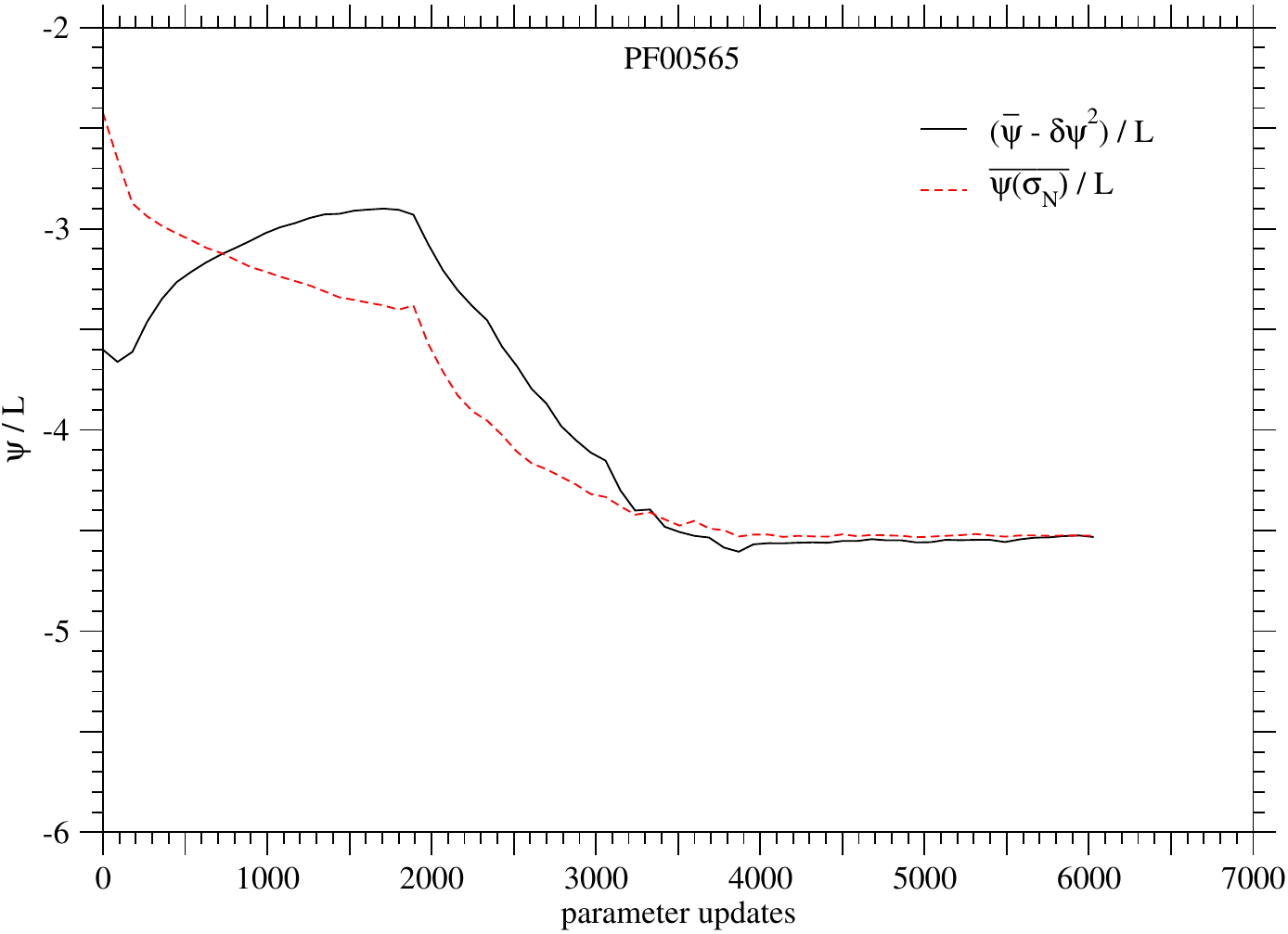}
}
% End of Figs_prt/PF00565/fig_prt.tex
}{% else
}
\caption
{
\TEXTBF{
For \PRT, the learning rate $\kappa(t)$, $D_2^{\text{KL}}$,
and $(\bar{\psi} - \delta\psi^2) / L$ and $\overline{\psi(\VEC{\sigma}_N)} / L$ at each step of 
the Boltzmann machine learning;}
to take account of the gauge invariance of interactions, the Ising gauge that satisfies \Eq{\ref{\EQ: Ising gauge}} is employed.
\label{\FIG: \PRT}
}
% End of Figs_prt/fig_prt_caption.tex
\end{figure*}
% End of Figs_prt/fig_prt_pf00565.tex

\renewcommand{\PRT}{PF00595}
\ifdefined\DIR
\renewcommand{\DIR}[1]{Figs_prt/#1}
\else
\newcommand{\DIR}[1]{Figs_prt/#1}
\fi
\ifdefined\SUBDIR
\renewcommand{\SUBDIR}[1]{\DIR{\PRT/#1}}
\else
\newcommand{\SUBDIR}[1]{\DIR{\PRT/#1}}
\fi
\ifthenelse{\equal{\PRT}{PF00595}}{%

\begin{figure*}[!ht]
\centerline{
\includegraphics[width=114mm,angle=0]{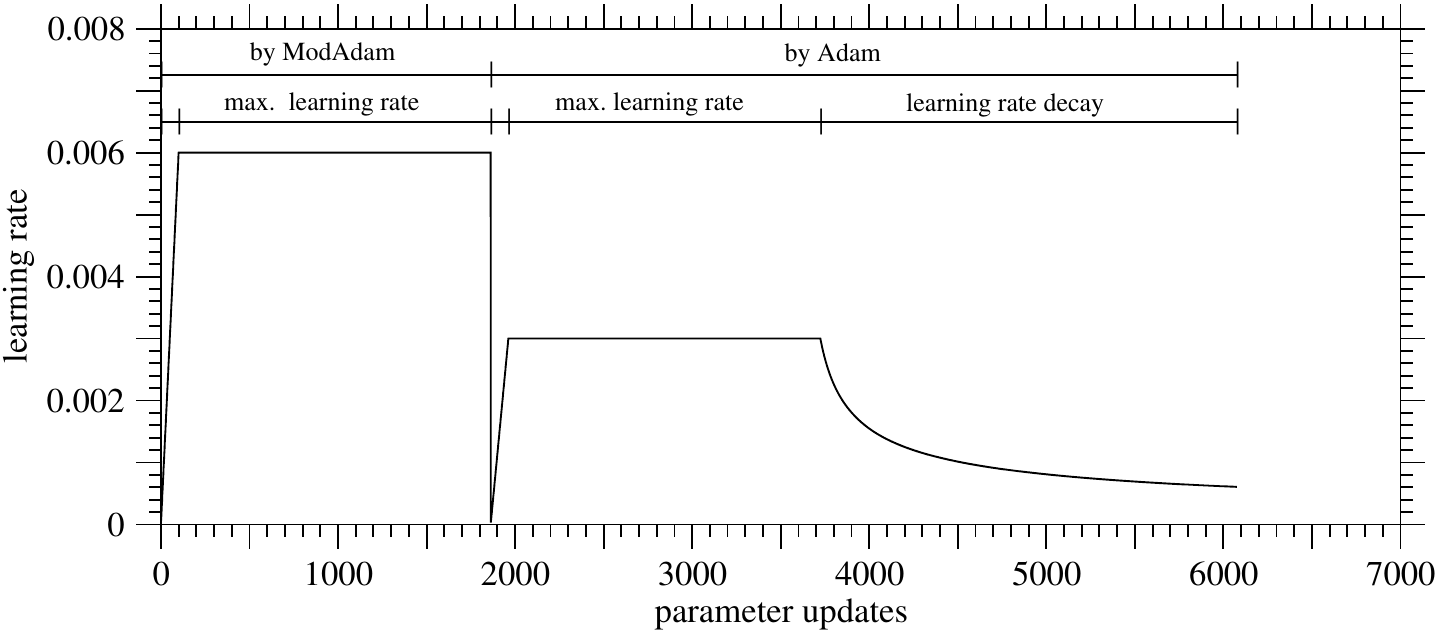}
}
\centerline{
\hspace*{3pt}
\includegraphics[width=112mm,angle=0]{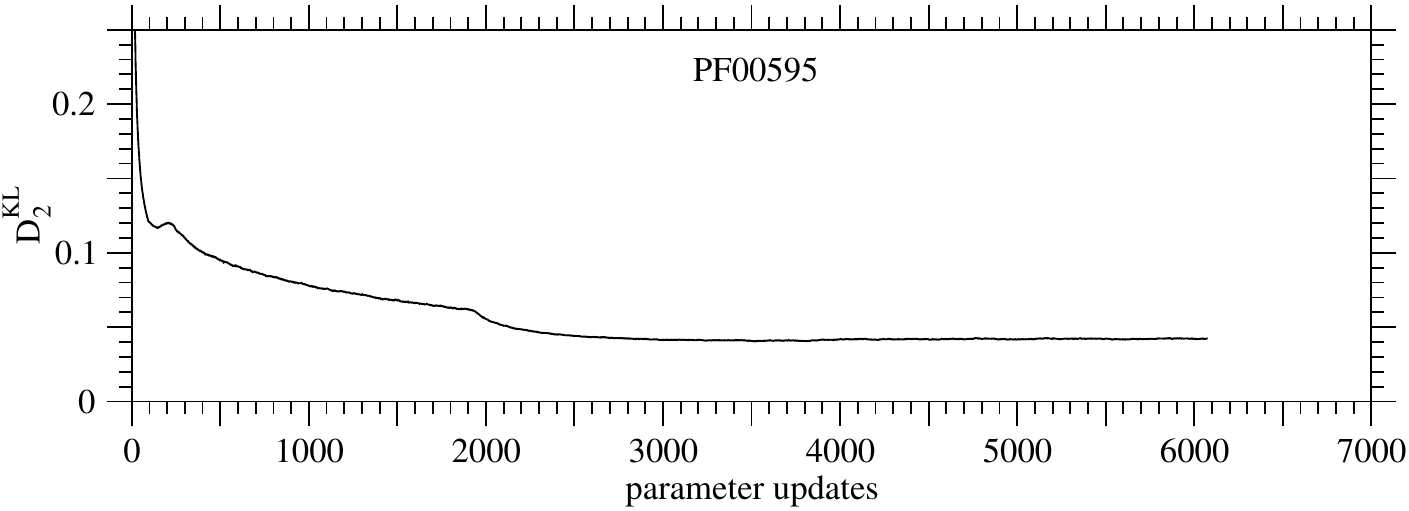}
}
\centerline{
\hspace*{7pt}
\includegraphics[width=110.5mm,angle=0]{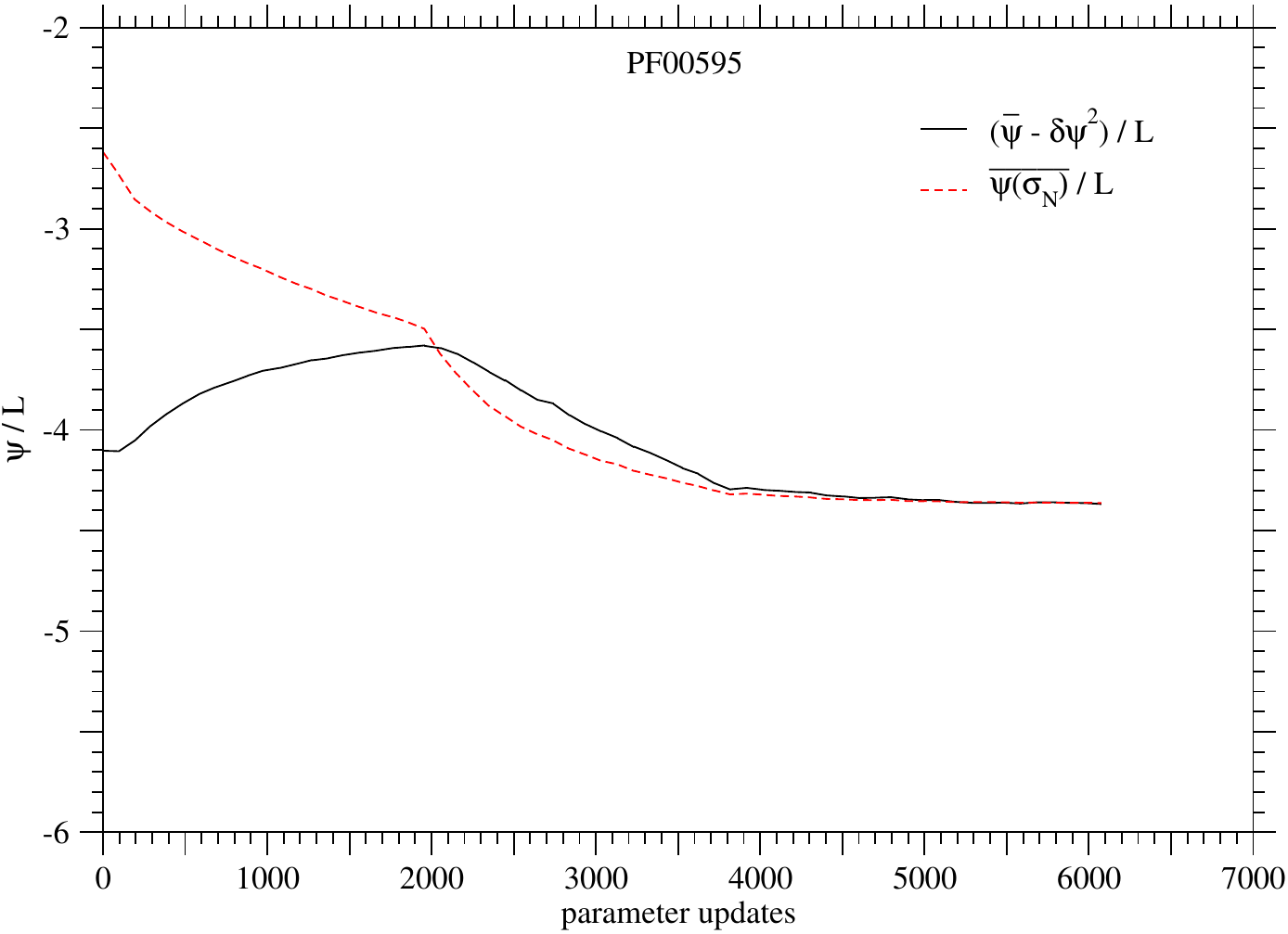}
}
% End of Figs_prt/PF00595/fig_prt.tex
}{% else
}
\caption
{
\TEXTBF{
For \PRT, the learning rate $\kappa(t)$, $D_2^{\text{KL}}$,
and $(\bar{\psi} - \delta\psi^2) / L$ and $\overline{\psi(\VEC{\sigma}_N)} / L$ at each step of 
the Boltzmann machine learning;}
to take account of the gauge invariance of interactions, the Ising gauge that satisfies \Eq{\ref{\EQ: Ising gauge}} is employed.
\label{\FIG: \PRT}
}
% End of Figs_prt/fig_prt_caption.tex
\end{figure*}
% End of Figs_prt/fig_prt_pf00595.tex

\renewcommand{\PRT}{PF00887}
\ifdefined\DIR
\renewcommand{\DIR}[1]{Figs_prt/#1}
\else
\newcommand{\DIR}[1]{Figs_prt/#1}
\fi
\ifdefined\SUBDIR
\renewcommand{\SUBDIR}[1]{\DIR{\PRT/#1}}
\else
\newcommand{\SUBDIR}[1]{\DIR{\PRT/#1}}
\fi
\ifthenelse{\equal{\PRT}{PF00887}}{%

\begin{figure*}[!ht]
\centerline{
\includegraphics[width=114mm,angle=0]{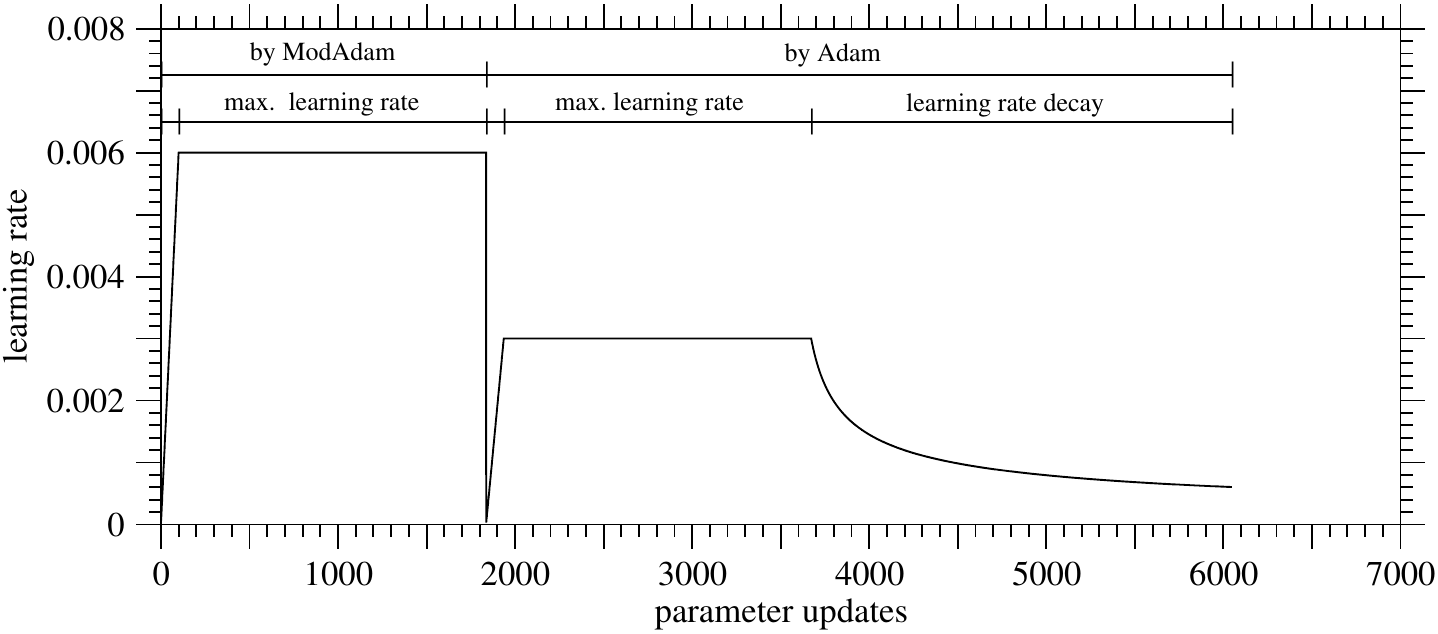}
}
\centerline{
\hspace*{3pt}
\includegraphics[width=112mm,angle=0]{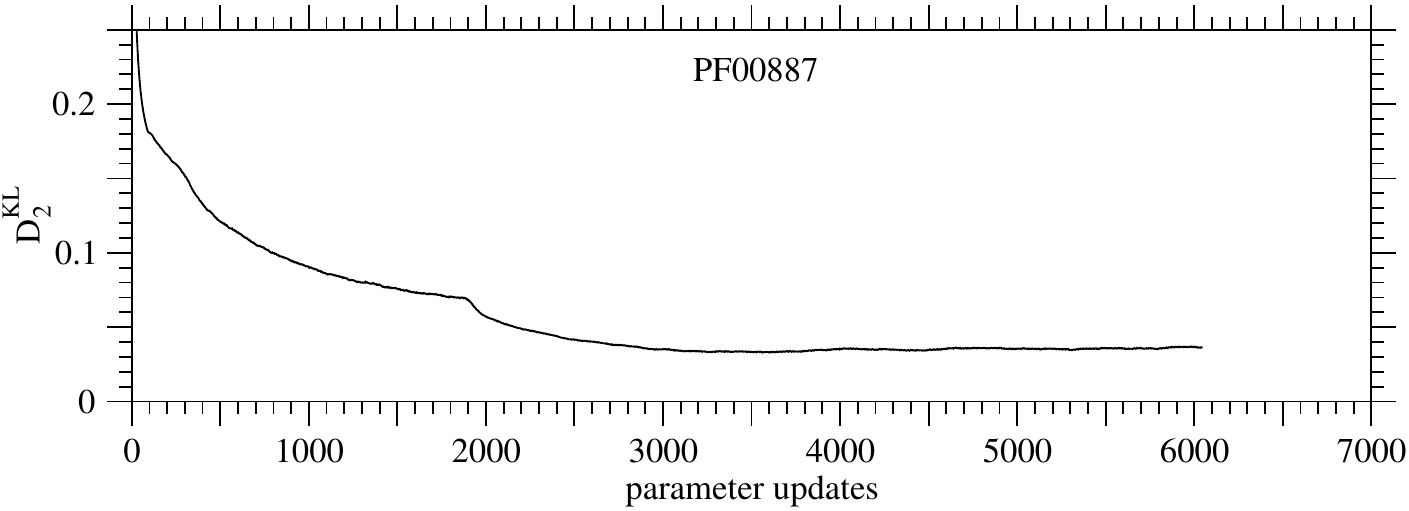}
}
\centerline{
\hspace*{7pt}
\includegraphics[width=110.5mm,angle=0]{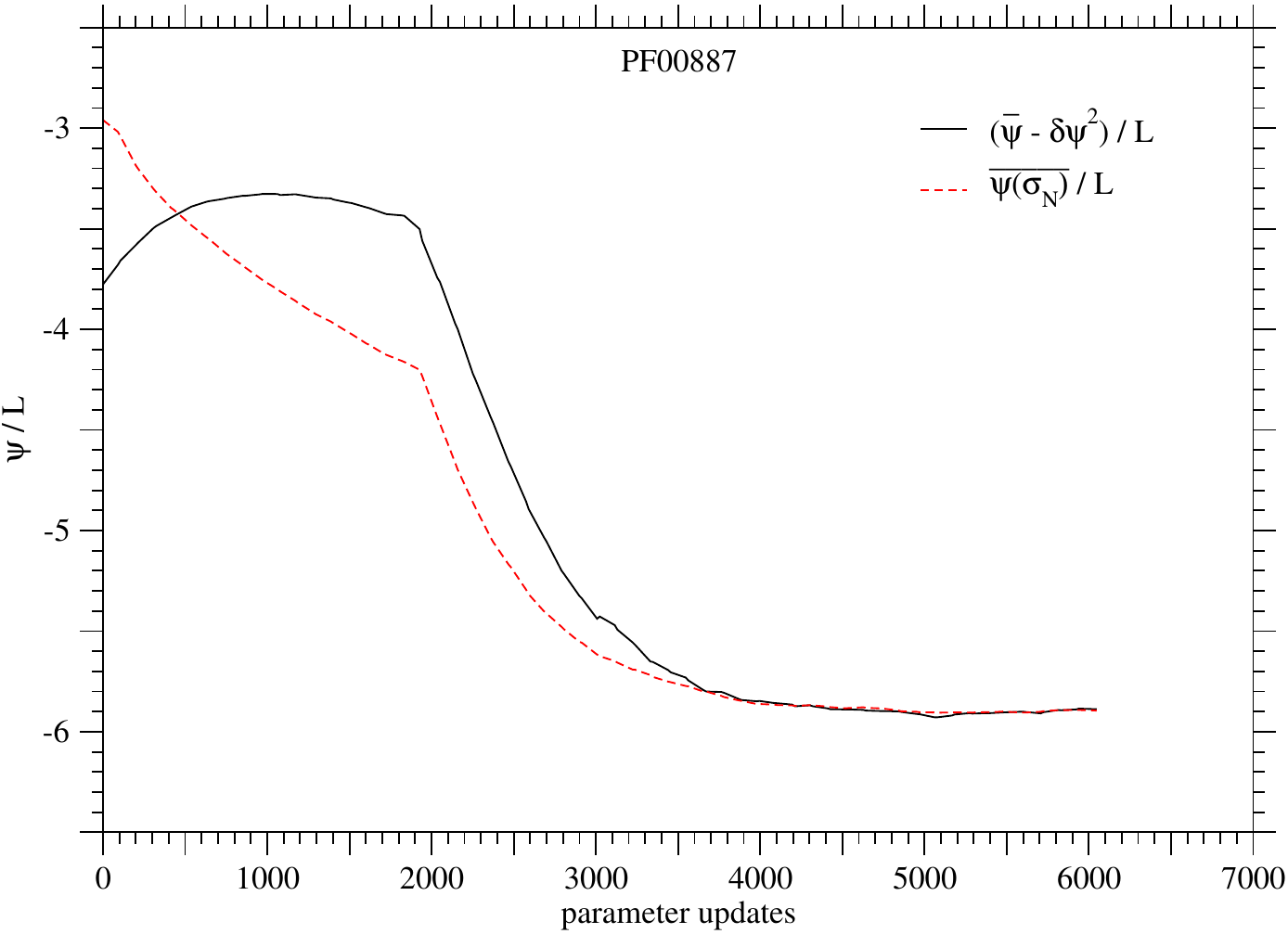}
}
% End of Figs_prt/PF00887/fig_prt.tex
}{% else
}
\caption
{
\TEXTBF{
For \PRT, the learning rate $\kappa(t)$, $D_2^{\text{KL}}$,
and $(\bar{\psi} - \delta\psi^2) / L$ and $\overline{\psi(\VEC{\sigma}_N)} / L$ at each step of 
the Boltzmann machine learning;}
to take account of the gauge invariance of interactions, the Ising gauge that satisfies \Eq{\ref{\EQ: Ising gauge}} is employed.
\label{\FIG: \PRT}
}
% End of Figs_prt/fig_prt_caption.tex
\end{figure*}
% End of Figs_prt/fig_prt_pf00887.tex

\renewcommand{\PRT}{PF00959}
\ifdefined\DIR
\renewcommand{\DIR}[1]{Figs_prt/#1}
\else
\newcommand{\DIR}[1]{Figs_prt/#1}
\fi
\ifdefined\SUBDIR
\renewcommand{\SUBDIR}[1]{\DIR{\PRT/#1}}
\else
\newcommand{\SUBDIR}[1]{\DIR{\PRT/#1}}
\fi
\ifthenelse{\equal{\PRT}{PF00959}}{%

\begin{figure*}[!ht]
\centerline{
\includegraphics[width=114mm,angle=0]{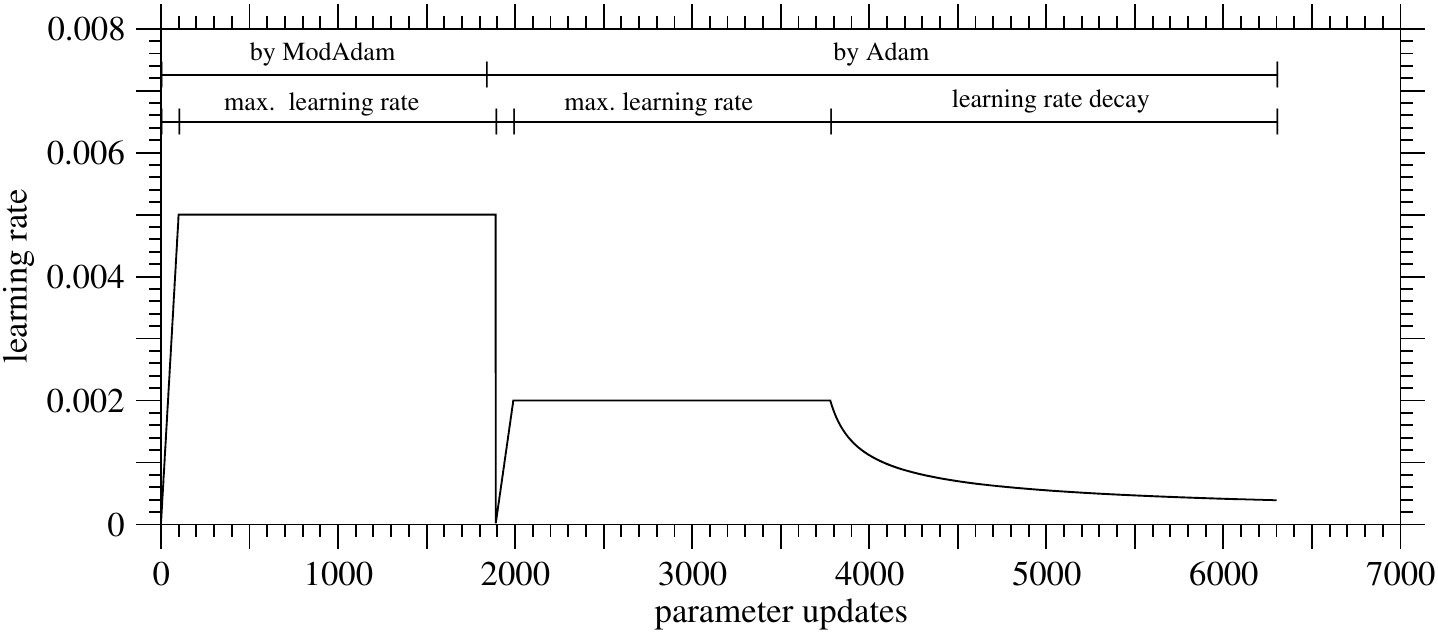}
}
\centerline{
\hspace*{3pt}
\includegraphics[width=112mm,angle=0]{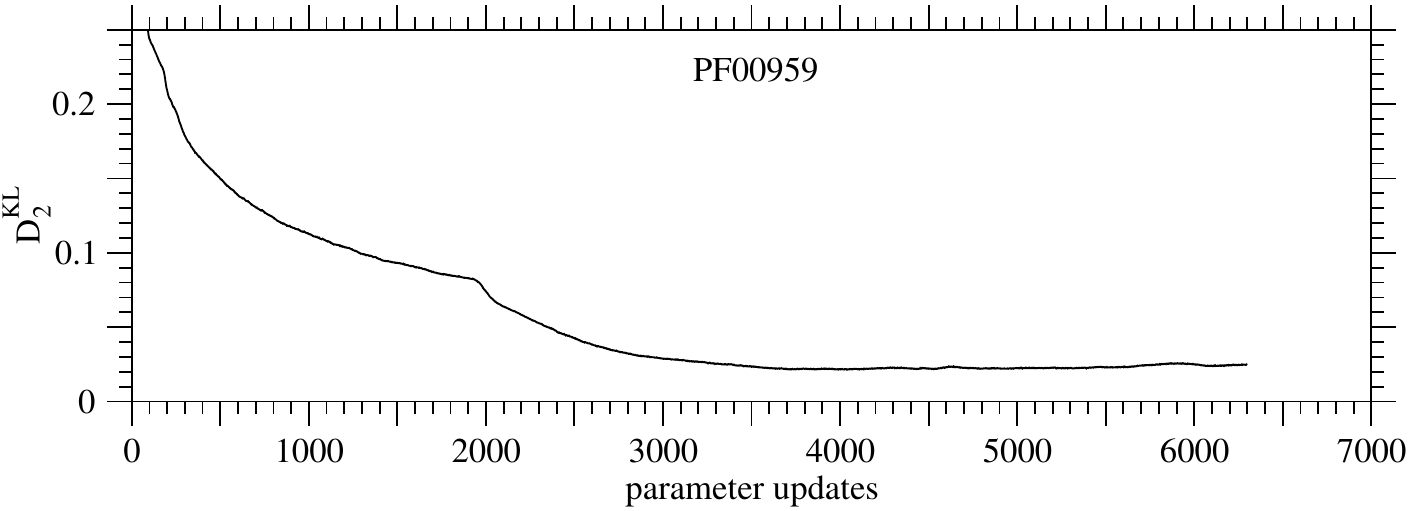}
}
\centerline{
\hspace*{7pt}
\includegraphics[width=110.5mm,angle=0]{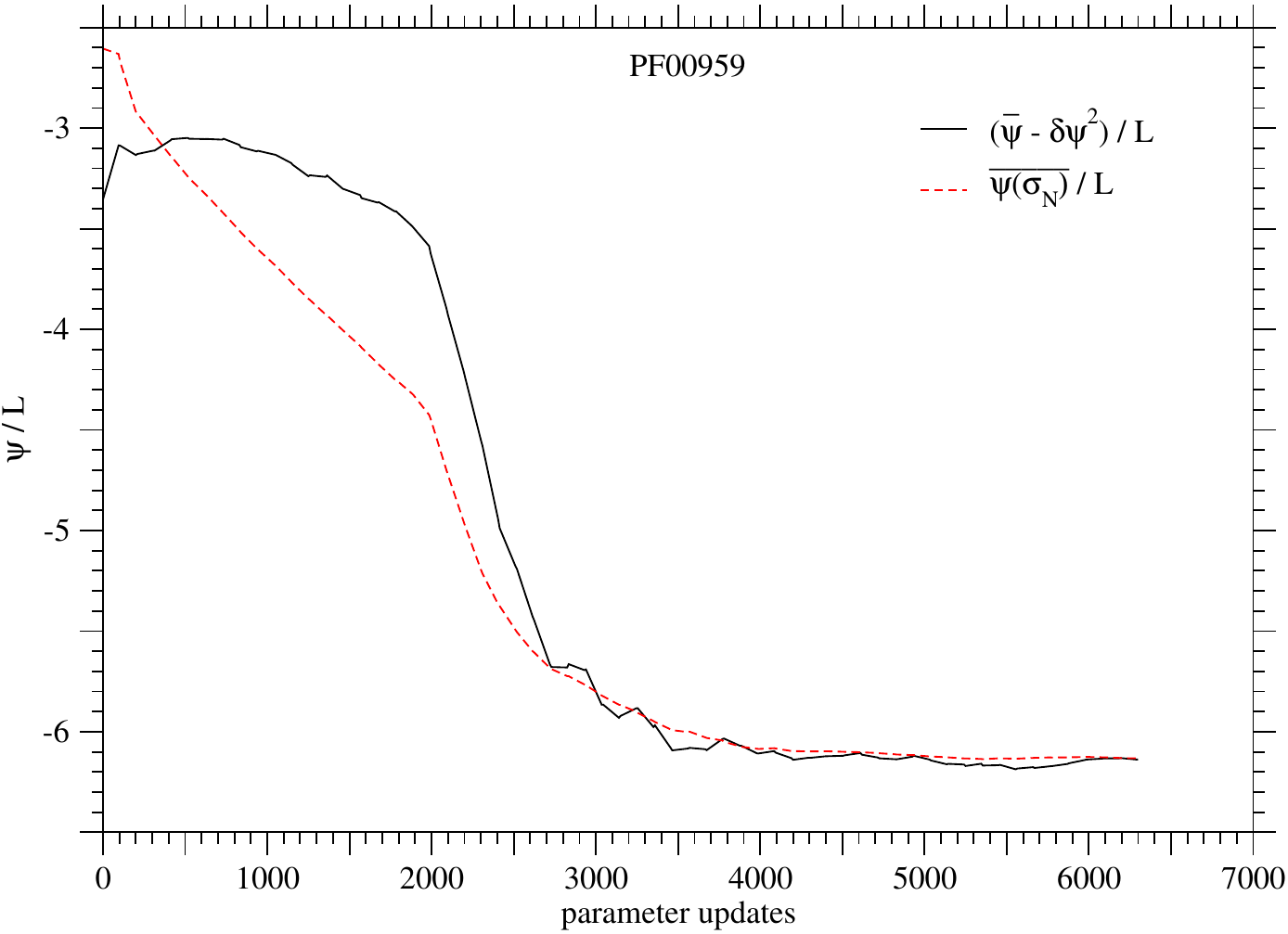}
}
% End of Figs_prt/PF00959/fig_prt.tex
}{% else
}
\caption
{
\TEXTBF{
For \PRT, the learning rate $\kappa(t)$, $D_2^{\text{KL}}$,
and $(\bar{\psi} - \delta\psi^2) / L$ and $\overline{\psi(\VEC{\sigma}_N)} / L$ at each step of 
the Boltzmann machine learning;}
to take account of the gauge invariance of interactions, the Ising gauge that satisfies \Eq{\ref{\EQ: Ising gauge}} is employed.
\label{\FIG: \PRT}
}
% End of Figs_prt/fig_prt_caption.tex
\end{figure*}
% End of Figs_prt/fig_prt_pf00959.tex

% End of Figs_prt/figs_prt.tex

}% SUPPLEMENT
}% SkipSupplToMerge
% End of tab-and-fig.tex

}{

}% TCBB

}% SkipSupplToMerge
}% SUPPLEMENT
% End of contents_supplement.tex

%%%%%%%%%%
}% SkipSupplToMerge
}% SUPPLEMENT

\end{document}